# Searching for magnetically hard monoborides (and finding a few): A first-principles investigation

Justyn Snarski-Adamski* and Mirosław Werwiński
*Institute of Molecular Physics, Polish Academy of Sciences,*
*M. Smoluchowskiego 17, 60-179 Poznań, Poland*

New hard magnetic materials with zero or low rare earth content are in demand due to the high prices of the rare earth metals. Among the candidates for such materials, we consider MnB, FeB and their alloys, because previous experiments suggest that FeB has a relatively high magnetic hardness of about 0.83 at room temperature. Using first-principles calculations, we examine the full range of alloys from CrB, through MnB, FeB, to CoB. Furthermore, we consider alloys of MnB and FeB with substitutions of $3d$, $4d$ and $5d$ transition metals. For the above ninety compositions, we determine magnetic moment, magnetocrystalline anisotropy energy and magnetic hardness.

For the alloys with transition metals we calculated also formation energies and Curie temperatures. For (Fe-Co)B alloys, the calculated values of magnetic hardness exceed five, which is an exceptionally high. While these values are inflated by the virtual crystal approximation used, we still expect actual magnetic hardnesses well above unity. Furthermore, we classify considered MnB alloys substituted with transition metals as magnetically soft or semi-hard and FeB alloys with Sc, Ti, V, Zr, Nb, Mo, Hf, Ta, or W as magnetically hard (with magnetic hardness exceeding unity).

## I. INTRODUCTION

In this theoretical work, we will consider the application of monoborides as magnetically hard materials. To this end, we will examine the full range of alloys from CrB, through MnB, FeB, to CoB, and also consider the complete series of MnB and FeB alloys with $3d$, $4d$, and $5d$ element substitutions. One of the main motivations for the search for magnetically hard monoborides is the promising experimental results for FeB (magnetic hardness around 0.83) [1, 2] and for half-borides [3, 4].

The basic magnetic properties of transition metal monoborides in a series passing through the elements V, Cr, Mn, Fe, Co, and Ni (with increasing atomic number) have been known since at least 1962 [5, 6], while the more recent advances in transition metal monoborides are summarized in Refs. [7] and [8]. MnB received significant attention between 2016 and 2023 [9–15]. This led to characterization of its lesser-known phases, such as $\alpha$ phase (CrB-type, s.g. $Cmcm$) [15] as well as $\alpha'$ phase (stacking-fault dominated CrB-like variant, s.g. $Cmcm$) [13]. In addition, the well-known $\beta$ phase (FeB-type, orthorhombic, s.g. $Pnma$) was also reexamined [9–11, 13, 15]. Moreover, the theoretical determination of the most energetically stable phases of FeB [16] and first-principles calculations for CrB [17] suggest the possibility of obtaining also a tetragonal MnB phase (Pearson symbol $tI16$). Conventionally, MnB is assumed to have a low-temperature mixed phase of CrB + FeB and a high-temperature FeB-type phase. [18].

In this work, we focus on the high-temperature phase $\beta$-MnB. It is a ferromagnetic phase with the magnetic moment equal to 1.9 $\mu_\mathrm{B}\,\mathrm{f.u.}^{-1}$ [15] (the highest value among all transition metal monoborides), saturation magnetization of 156 $\mathrm{Am^2\,kg^{-1}}$ (at 10 K) [15], magnetization remanence of 9.0 $\mathrm{Am^2\,kg^{-1}}$ [12], relatively high Curie temperature of 568 K [15], low coercive field of 15.9 Oe, classifying MnB as a magnetically soft material [10], enormously high Vickers hardness of 15.7 GPa [10], and a large magnetocaloric effect driven by anisotropic magnetoelastic coupling [11].

Similar to MnB, the FeB crystallizes in a low-temperature $\alpha$-phase (CrB-type) [11, 18, 19], which detailed structural properties have been recently discussed [20] along with a structural analysis of a high-temperature $\beta$ phase (orthorhombic, s.g. $Pnma$). As we already mentioned, for FeB theory also suggests an energetically stable tetragonal phase ($tI16$) [16]. For the $\beta$-FeB phase, the Curie temperature is relatively high at around 590 K [21], the magnetic moment (at 4 K) is equal to 1.1 $\mu_\mathrm{B}\,\mathrm{f.u.}^{-1}$ [22], and the effective magnetic anisotropy constant $K_1^*$ (for orthorhombic system) is equal to 0.4 $\mathrm{MJ\,m^{-3}}$. This leads to a magnetic hardness of 0.83 and just slightly fails to meet criterion for magnetically hard materials ($\kappa = 1$). [2]. However, the coercivity of FeB is low, similar to that of MnB, and equal to 10 Oe at 290 K [21]. The range of monoborides we consider in this work is limited on both sides by CrB (s.g. $Cmcm$) and CoB (s.g. $Pnma$) [17, 23].

One method of tuning the properties of monoborides is nanostructurization [7, 8, 21]. Whereby, particularly interesting form of monoborides are nanosheets [24], as experimentally obtained in two-dimensional CrB [25], MnB [26] and CoB [27] systems. In the case of MnB nanosheets, the saturation magnetization is equal to 15.5 $\mathrm{Am^2\,kg^{-1}}$ ($\mathrm{emu\,g^{-1}}$) at 300 K, and the Curie temperature of 586 K is the highest among all known two-dimensional ferromagnets [25]. Recent experimental efforts on monoboride nanosheets are complemented by theoretical work on two-dimensional CrB [28, 29], MnB [26, 29–31], FeB [28, 29, 32], and CoB [27, 32], which predicts both the magnetic properties of these systems and their dynamical stability. The calculations for magnetic monoboride nanosheets complement earlier first-principles studies on bulk CrB [17, 33, 34], MnB [10, 35, 36], and FeB [16, 37], as well as work comparing various bulk magnetic monoborides [11, 38–40].

* Corresponding author: Justyn Snarski-Adamski
justyn.snarski-adamski@ifmpan.poznan.pl

Transition metal borides are involved in energy-related electrocatalysis [8]. FeB, CoB, and NiB are being investigated as thermoelectrics [41], and nanoscale boron powders are being considered for hyperthermia and boron neutron capture therapy [21]. Two-dimensional transition metal borides are being investigated for magnetic information storage [28, 42]. CrB-based devices have been considered as sensors or adsorbents [43], and CrB nanosheets as promising anode materials in lithium-ion batteries [25]. MnB and its alloys exhibit a large magnetocaloric effect [9, 11, 15, 44] and very high Vickers hardness (15.6 GPa), combined with ferromagnetic properties, make MnB a promising mechanically hard magnetic material [10]. In addition, MnB is suggested for magnetic hyperthermia [15] and improves the hard magnetic properties of a combined hard / soft magnetic system (98% MnAl-C / 2% MnB) [12]. On the other hand, amorphous Co-B is considered as a catalyst [45] and CoB nanosheets as an electrocatalyst for metal-air batteries [27]. Finally, an example of a high-entropy pseudo-monoboride is $(Cr_{0.2}Mn_{0.2}Fe_{0.2}Co_{0.2}Mo_{0.2})B$ [46].

The long list above does not include applications of magnetic monoborides as magnetically hard materials, most likely due to the fact that coercivity measurements indicate that MnB and FeB are magnetically soft [10, 21]. However, given the arguments we are about to present, we decided to theoretically explore the possibility of using alloys of MnB and FeB as magnetically hard materials. To this end, we choose to calculate from first principles the magnetic hardness values for about ninety chemical compositions, which we hope will allow us to propose promising candidates for permanent magnets or discredit the idea. The following arguments led us to consider the use of monoboride alloys in permanent magnets: (1) reasonably high Curie temperatures, above 540 K for MnB and FeB, and above 800 K for (Mn,Fe)B alloys [1, 10, 21], (2) reasonably high saturation magnetization of MnB and FeB [1, 10, 21], (3) orthorhombic crystal system of MnB and FeB, which can provide a good starting point for the development of magnetic hardness, (4) relatively high experimental effective magnetocrystalline anisotropy constant of FeB monocrystal ($K_1^* = 0.4$ MJ m$^{-3}$) [2], (5) relatively high magnetic hardness of FeB which we estimated as 0.83 from the experimental $K_1^*$, magnetization and lattice parameters [1, 2], (6) elevated magnetic hardness of annealed $\alpha$-FeB [21] and FeB nanoparticles [20], (7) semi-hard magnetic properties determined for some of $(Fe,Co)_5PB_2$ [47–49], $(Fe,Co)_3B$ [50], and $(Fe,Co)_2B$ [3, 4, 51] borides, and last but important (8) excellent hard magnetic properties of another transition metal boride, Nd-Fe-B, one of the most popular permanent magnets today.

The results presented here provide a systematic description of the rudamental magnetic properties of doped monoborides, which can be used in many of the numerous potential applications listed above. However, especially in the context of volatile neodymium prices observed between 2011 and 2024 [52], it is important to search for new permanent magnets that contain little or no rare-earth elements [53, 54].

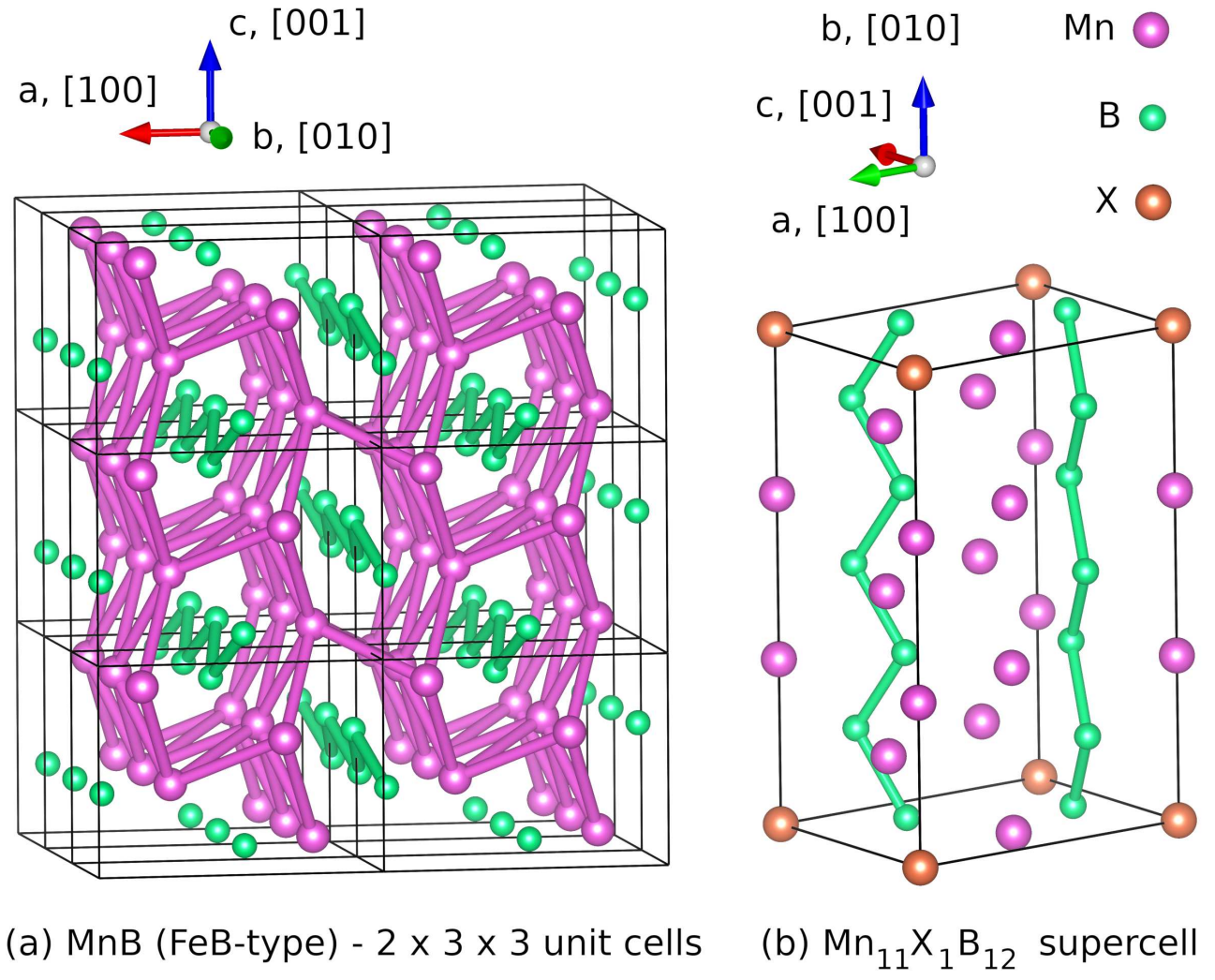

FIG. 1. (a) Crystal structure of $\beta$-MnB (FeB-type; space group $Pnma$). The lattice parameters of a single unit cell are $a = 5.6377$ Å, $b = 2.9945$ Å, and $c = 4.1795$ Å [10]. The unit cell is multiplied $2 \times 3 \times 3$ times along the main axes for better visualization. (b) Crystal structure of $Mn_{11}X_1B_{12}$ supercell (space group $P1m1$) with transition metal substitution X. The supercell was obtained by duplicating the MnB unit cell three times along the $b$-axis and substituting one of twelve Mn atoms.

## II. COMPUTATIONAL DETAILS

For density functional theory calculations, we used the full-potential local-orbital electronic structure code FPLO18.00-52 including relativistic effects in the full four-component formalism [55, 56]. Unless otherwise noted, we employed the generalized gradient approximation (GGA) in the form proposed by Perdew, Burke, and Ernzerhof (PBE) [57]. We choose a tetrahedron method for Brillouin zone integration. All calculations were fully converged with density criterion $10^{-6}$.

For MnB, we assumed a $Pnma$ space group (FeB-type) with experimental lattice parameters $a = 5.6377$ Å, $b = 2.9945$ Å, and $c = 4.1795$ Å, and initial Wyckoff positions from Ref. [10], see also Table I and Fig. 1(a). For FeB, we adopted the space group $Pnma$ with experimental lattice parameters $a = 5.4954$ Å, $b = 2.9408$ Å, and $c = 4.0477$ Å, and initial Wyckoff positions from Ref. [19]. We optimized the initial atomic positions and present our results in Table I as well. The Wyckoff positions predicted theoretically for MnB differ significantly from the experimental results from work of Ma *et al.* [10]. The calculated positions are instead very close to another ex-perimental result for MnB from work of Kalyon *et al.* [15] and to the results for isostructural FeB [19]. Regardless, we use the optimized atomic positions throughout our work.

TABLE I. Experimental and theoretical Wyckoff positions of MnB and FeB (s.g. *Pnma*). Atomic positions for B and M (M = Mn or Fe) are presented. The DFT results were calculated with FPLO18 code and the GGA-PBE exchange-correlation potential.

| | | Ref. | $x$(M) | $y$(M) | $z$(M) | $x$(B) | $y$(B) | $z$(B) |
|---|---|---|---|---|---|---|---|---|
| MnB | expt. | [10] | 0.1680 | 0.25 | 0.1244 | 0.1482 | 0.25 | 0.6375 |
| MnB | expt. | [15] | 0.1762 | 0.25 | 0.1206 | 0.0312 | 0.25 | 0.6133 |
| MnB | PBE | | 0.1746 | 0.25 | 0.1211 | 0.0330 | 0.25 | 0.6169 |
| FeB | expt. | [19] | 0.1771 | 0.25 | 0.1195 | 0.0332 | 0.25 | 0.6168 |
| FeB | PBE | | 0.1776 | 0.25 | 0.1199 | 0.0344 | 0.25 | 0.6200 |

For modeling the solid solutions in a range from CrB ($Z = 24$), through MnB ($Z = 25$), FeB ($Z = 26$), to CoB ($Z = 27$) we employed virtual crystal approximation (VCA). For the whole range of VCA compositions, we kept the space group *Pnma*. For MnB [10], FeB [19], and CoB [23] we used experimental lattice parameters and optimized atomic positions. For the CrB-MnB range, we set lattice parameters and Wyckoff positions extrapolated from the MnB/FeB pair. For other intermediate concentrations, we used interpolated values of lattice parameters and Wyckoff positions. To determine the magnetocrystalline anisotropy energy (MAE) within the VCA, we performed self-consistent scalar-relativistic calculations, followed by a single fully-relativistic iteration for each of three main crystallographic axes. We used $11 \times 20 \times 15$ k-mesh.

To determine the dependency of magnetocrystalline anisotropy energy on magnetic moment of FeB we used a fully-relativistic fixed-spin-moment method. Moreover, we used the local spin density approximation (LSDA) in the forms of von Barth and Hedin (BH) [58] and Perdew and Wang (PW92) [59] to analyze the effect of the form of exchange-correlation potential on the magnetic moment and MAE.

Based on $Mn_{11}X_1B_{12}$ and $Fe_{11}X_1B_{12}$ supercells, we studied the effect of substituting MnB and FeB with $3d$, $4d$, and $5d$ elements. The supercells are three-times repetitions in $b$ direction of *Pnma* unit cell, see Fig. 1(b). The Wyckoff positions of MnB and FeB unit cells were first optimized, assuming the force convergence criterion of $10^{-3}$ eV Å$^{-1}$. The lattice parameters of the supercells were kept fixed at $a = 5.6377$ Å, $b = 8.9835$ Å, and $c = 4.1795$ Å for $Mn_{11}X_1B_{12}$ [10] and $a = 5.4954$ Å, $b = 8.8224$ Å, and $c = 4.0477$ Å for $Fe_{11}X_1B_{12}$ [19]. The reduced space group of these structures was $P1m1$. For supercells, we choose an $8 \times 5 \times 11$ k-mesh. Drawings of the structures were made using the VESTA code [60].

To determine the magnetic hardness, we need to calculate the magnetocrystalline anisotropy energy. For uniaxial crystal systems, such as tetragonal and hexagonal, the MAE can be evaluated from the difference in energies determined for the axis along the unique crystallographic axis and the axis perpendicular to it. Here, since FeB and MnB crystallize in an orthorhombic system characterized by three different lattice parameters $a$, $b$, and $c$, the approach used for uniaxial systems cannot be directly applied. Usually in that case, only the magnetic easy axis is determined [36]. In this work, to quantify the MAE, we use the following approach. We calculate the fully relativistic energies for three quantization directions along the main crystallographic axes: [100] (along $a$), [010] (along $b$), and [001] (along $c$).

The lowest energy indicates the magnetic easy axis, and the difference between the middle energy ($E_2$) and the lowest energy ($E_1$) determines the MAE. As a result, all values of MAE are non-negative. One could also take the difference between the highest ($E_3$) and lowest ($E_1$) energy. However, we opted for the first option, interpreting MAE as a parameter that determines the uniaxial preference for magnetization direction. To complement this parameter, we calculate also the energy difference between the highest ($E_3$) and the middle ($E_2$) of the three energies, which we denote as magnetocrystalline anisotropy energy $DE_{32}$. A similar approach was used by Zhdanova *et al.* [2] to experimentally determine the effective magnetocrystalline anisotropy constants $K_1^* = K_1 - K_2$ and $K_2^* = K_3 - K_2$ of orthorhombic FeB, where $K_1$, $K_2$, and $K_3$ are the three anisotropy constants determined in the three main crystallographic planes and arranged in descending order.

When discussing applications for permanent magnets, magnetic hardness is another crucial metric [54]. It can be expressed as:

$$\kappa = \sqrt{\frac{|K|}{\mu_0 M_S^2}}, \tag{1}$$

where $K$ stands for the magnetocrystalline anisotropy constant, $M_S$ denotes the saturation magnetization, and $\mu_0$ the vacuum permeability. In order to determine the theoretical value of $\kappa$, we assume that the calculated MAE is equal to the anisotropy constant $K$. To estimate the value of $M_S$ we use the total magnetic moment and the unit cell volume.

Starting from the magnetic force theorem [61, 62], we determined the contributions from each **k**-point to the MAE using the equation:

$$\begin{aligned}\text{MAE} &= E(\theta = 90^\circ) - E(\theta = 0^\circ) = \\ &= \sum_{\text{occ'}} \epsilon_i(\theta = 90^\circ) - \sum_{\text{occ''}} \epsilon_i(\theta = 0^\circ),\end{aligned} \tag{2}$$

where $\theta$ is the angle between the direction of magnetization and the selected axis, $E(\theta)$ is the total energy for a specific direction; and $\epsilon_i$ is the band energy of the $i$ state, see Ref. [48].

Furthermore, we determined the formation energies of $Fe_{11}X_1B_{12}$ compositions from the formula:

$$E_f = E_{Fe_{11}X_1B_{12}} - 11E_{Fe} - E_X - 12E_B, \tag{3}$$

where $E_{Fe_{11}X_1B_{12}}$, $E_{Fe}$, $E_B$, and $E_X$ are the total energies of the $E_{Fe_{11}X_1B_{12}}$ supercell and the crystals of iron, boron, and transition metal (X), respectively. In order to determine the total energies, we prepared and optimized the geometry of 3$d$, 4$d$, and 5$d$ elemental bulk materials. Formation energies of $Mn_{11}X_1B_{12}$ compositions we established in an analogous manner.

Moreover, using the mean-field theory, we calculated the Curie temperature ($T_C^{MFT}$) of $Fe_{0.917}X_{0.083}B$ alloys with 3$d$, 4$d$, 5$d$ transition-metal elements (X) [63, 64]. We used the equation:

$$k_B T_C^{MFT} = \frac{2}{3}(E_{DLM} - E_{FM}), \quad (4)$$

where $E_{DLM}$ and $E_{FM}$ are the total energies of paramagnetic and ferromagnetic solutions, and $k_B$ is Boltzmann constant. We model the paramagnetic configuration using the disordered local moment (DLM) approach [65] with the coherent potential approximation (CPA) [66]. The scalar-relativistic DLM-CPA calculations were performed using the FPLO5 code, which is the latest public version of the FPLO code including CPA. Unfortunately, it does not have the PBE-GGA implemented, thus the DLM calculations were performed with the PW92-LDA exchange-correlation potential [59].

In this work, we have determined the magnetic properties for temperature of 0 K (ground state), while permanent magnets usually operate at room or higher temperature. While in this situation the calculated magnetic moments can be reasonably treated as upper limits, the issue of MAE's dependence on temperature is more complex. For example, it has been experimentally shown for magnetic halfborides $(Fe\text{-}Co)_2B$, that magnetocrystalline anisotropy constant ($K_1$) depends on temperature in a nonlinear manner (including a change of sign) [4]. However, a similarity was observed by us between the concentration dependence and temperature dependence of $K_1$. Thus, we assume that the highest MAEs among the values determined at 0 K for the full range of the concentrations will at the same time represent a limit in the temperature dependence of MAEs.

Furthermore, the MAE values determined for alloys in the VCA are often overestimated [3]. However, calculations for alloys modeled by the supercell method (without VCA) yield similar and in some cases even higher MAE and $\kappa$ values than those determined by VCA for (Mn-Fe)B compositions, which makes our VCA results more reliable. Another constraint of the calculation method we employed is the accuracy of the GGA. Therefore, we will discuss the effect of the form of the exchange-correlation potential on the obtained MAE values [67]. Finally, the other known limitation is lack of disorder in supercells due to their limited size, which can also affect the determined MAE [51, 67].

## III. RESULTS AND DISCUSSION

Problems with the availability and prices of rare earth ores [52] are stimulating the search for new permanent magnets with zero or low rare earth content [3, 4, 47, 51, 53, 54]. As part of this investigation, here, we consider monoboride alloys ranging from CrB ($Z = 24$), through MnB (25) and FeB (26), to CoB (27), or in other words, concentration ranges: $Cr_{1-x}Mn_xB$, $Mn_{1-x}Fe_xB$, and $Fe_{1-x}Co_xB$. Furthermore, we study MnB and FeB alloys with substitutions of thirty different transition metal elements.

In the *Introduction* we said that high-temperature phase of both MnB and FeB is orthorhombic phase with space group $Pnma$ [18]. Despite the experimentally confirmed distinct low-temperature phases, and theoretical predictions of a stable tetragonal phase in the ground state, in this computational work we consider MnB, FeB, and their alloys in high-temperature phase $Pnma$. Our choice is dictated by the fact that the orthorhombic phase characterizes the studied systems at room temperature and above – the range in which we expect their practical applications as permanent magnets.

### A. (Cr-Mn)B, (Mn-Fe)B, and (Fe-Co)B alloys from virtual crystal approximation

From the perspective of application as permanent magnets, the three most important intrinsic properties of a magnetic material are Curie temperature, saturation magnetization and magnetocrystalline anisotropy. Below, we will summarize all three for magnetic monoborides, with the last two quantities expressed in a form of magnetic moment and magnetocrystalline anisotropy energy. Although the concentration dependence of Curie temperature and magnetic moments have long been known, our calculations complement the picture, providing the values of magnetocrystalline anisotropy energy and magnetic hardness.

TABLE II. Calculated magnetic properties of characteristic compositions in the MnB-FeB-CoB range (s.g. $Pnma$). The magnetic easy axis, spin and orbital magnetic moment [$m_s$ and $m_l$ ($\mu_B$ atom$^{-1}$)] on boron (B) and virtual atom of transition metal (M), magnetocrystalline anisotropy energy [MAE (MJ m$^{-3}$)], and magnetic hardness [$\kappa$] are presented. Orbital magnetic moments are given for the corresponding magnetic easy axes. Calculations were performed with FPLO18-PBE in the virtual crystal approximation (VCA).

| Composition | Axis | $m_s$(M) | $m_l$(M) | $m_s$(B) | $m_l$(B) | MAE | $\kappa$ |
|---|---|---|---|---|---|---|---|
| MnB | 010 | 2.18 | 0.023 | -0.20 | 0.001 | 0.45 | 0.57 |
| $Mn_{0.55}Fe_{0.45}B$ | 100 | 1.92 | 0.026 | -0.18 | 0.000 | 0.77 | 0.83 |
| $Mn_{0.15}Fe_{0.85}B$ | 010 | 1.53 | 0.028 | -0.14 | 0.001 | 0.69 | 0.95 |
| FeB | 010 | 1.33 | 0.025 | -0.13 | 0.001 | 0.14 | 0.49 |
| $Fe_{0.5}Co_{0.5}B$ | 001 | 0.61 | 0.047 | -0.06 | 0.001 | 2.87 | 5.09 |

The first systematic measurements of the properties of CrB-MnB and MnB-FeB alloys were made by Cadeville and Daniel in 1966 [6]. The research was then continued by Kanaizuka [1, 22]. It was found that CrB ($Z = 24$ for Cr) is paramagnetic and at intermediate concentration (Cr-Mn)B alloys gain ferromagnetic ordering. The Curie temperature increases with Mn concentration, and for MnB it is about 575 K, see Fig. 2(a) [1, 6]. For (Mn-Fe)B alloys, Curie temperature grows above 800 K for $Mn_{0.4}Fe_{0.6}B$. For (Fe-Co)B compositions, Curie temperature decreases with Co concentration to zero, and CoB itself is paramagnetic [68, 69]. From the considered perspective of application for permanent magnets, Curie temperature values well above room temperature are required. Hence, MnB-FeB alloys located around the Curie temperature maximum would be best suited.

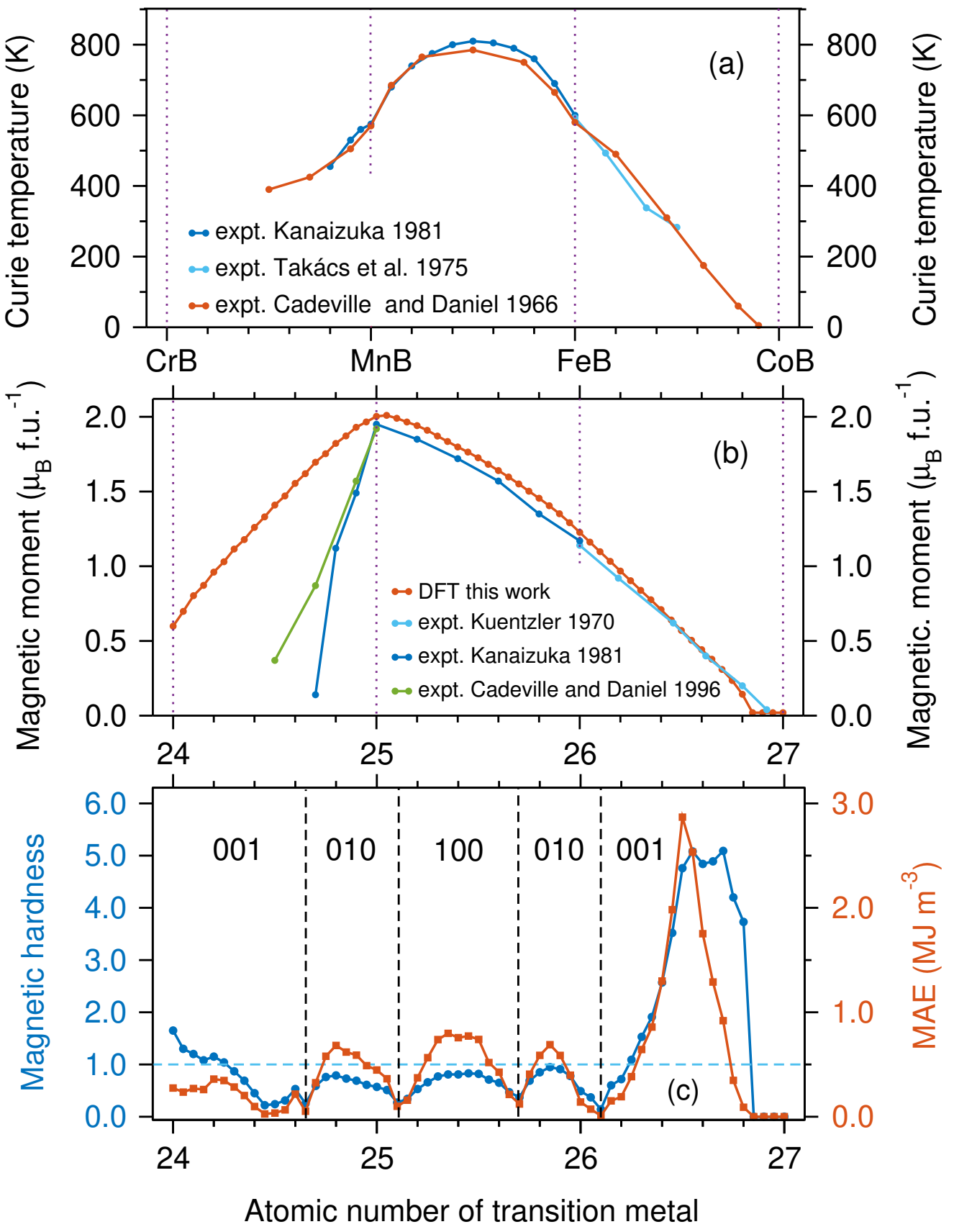

FIG. 2. Intrinsic properties critical for permanent magnet applications obtained for the concentration range from CrB, through MnB and FeB, to CoB. (a) Experimental Curie temperatures taken from the literature. (b) Calculated total spin magnetic moments as a function of the atomic number of the transition metal element, together with the corresponding experimental total magnetic moments from the literature. (c) Calculated magnetocrystalline anisotropy energy (MAE) and magnetic hardness as a function of atomic number. Vertical dashed lines separate alloys with different axes of easy magnetization. Calculations were performed with the virtual crystal approximation (VCA) using the FPLO18 code with PBE exchange-correlation potential. Experimental data are taken from Refs. [1, 6, 68, 69].

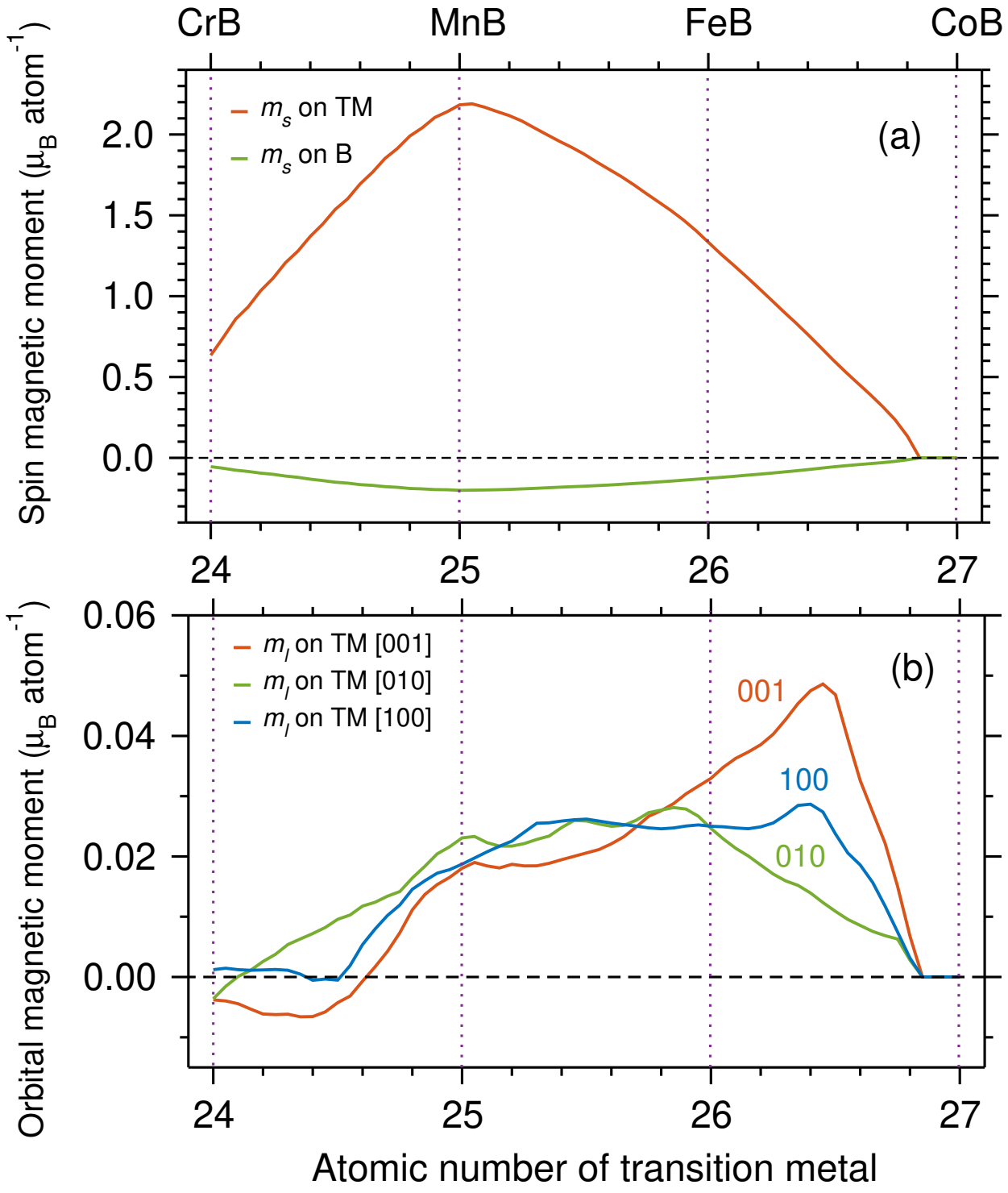

FIG. 3. Spin and orbital magnetic moments ($m_s$ and $m_l$) on transition metal (TM) and boron sites as a function of the atomic number of the transition metal element, calculated for concentration range from CrB, through MnB and FeB, to CoB. The calculations were performed with the virtual crystal approximation (VCA) using the FPLO18 code with PBE exchange-correlation potential.

Experimental studies of the concentration dependence of the magnetic moment in the CrB-MnB-FeB-CoB range reveal the absence of an ordered magnetic moment on the Cr-rich side and the Co-rich side, see Fig. 2(b) [1, 5, 6, 68, 69]. MnB has the maximum value of the magnetic moment of about 1.9 $\mu_B$. Passing from MnB through FeB to CoB, we observe its linear decrease. The experimental dependence of magnetic moments on concentration is well reflected in the previous DFT results for the MnB-FeB [9, 70, 71] and FeB-CoB [72, 73] solid solutions. Our results for the MnB-FeB-CoB range also well reproduce the experimental dependence of the magnetic moment on concentration. Although for the CrB-MnB range, the calculated values differ from the experimental ones. This is most likely due to the adoption of an identical *Pnma* space group for all VCA concentrations, while a *Cmcm* space group (MoB-type structure) is experimentally observed for CrB and nearby [17, 18, 22]. At the other end of the range considered, we observe a collapse of the magnetic moment above 80% of the Co concentration in (Fe-Co)B alloys, similar to previous experimental [6, 68] and theoretical [72, 73] studies.

Before moving on to the results on magnetic hardness, we would like to discuss the spin and orbital magnetic moments on individual atoms, see Fig. 3. The largest contributions to the total magnetic moment come from the spin magnetic moment on the transition metal site and from a smaller induced opposite spin moment on boron. For these spin moments, we have shown values for one representative direction [010], since the other two cases are indistinguishable. For the orbital moments on transition metal site, we have plotted values calculated in three directions. The differences observed for orbital moments determined in various directions correlate with the magnetocrystalline anisotropy constant, as given by Bruno's formula [74]. In our case, the large differences in orbital moments for FeB-CoB alloys correlate with the high magnetic anisotropy energies shown in Fig. 2. The values of the orbital magnetic moments calculated in the PBE approximation are about twice as small as the experimental ones, as is the case with bcc Fe [75].

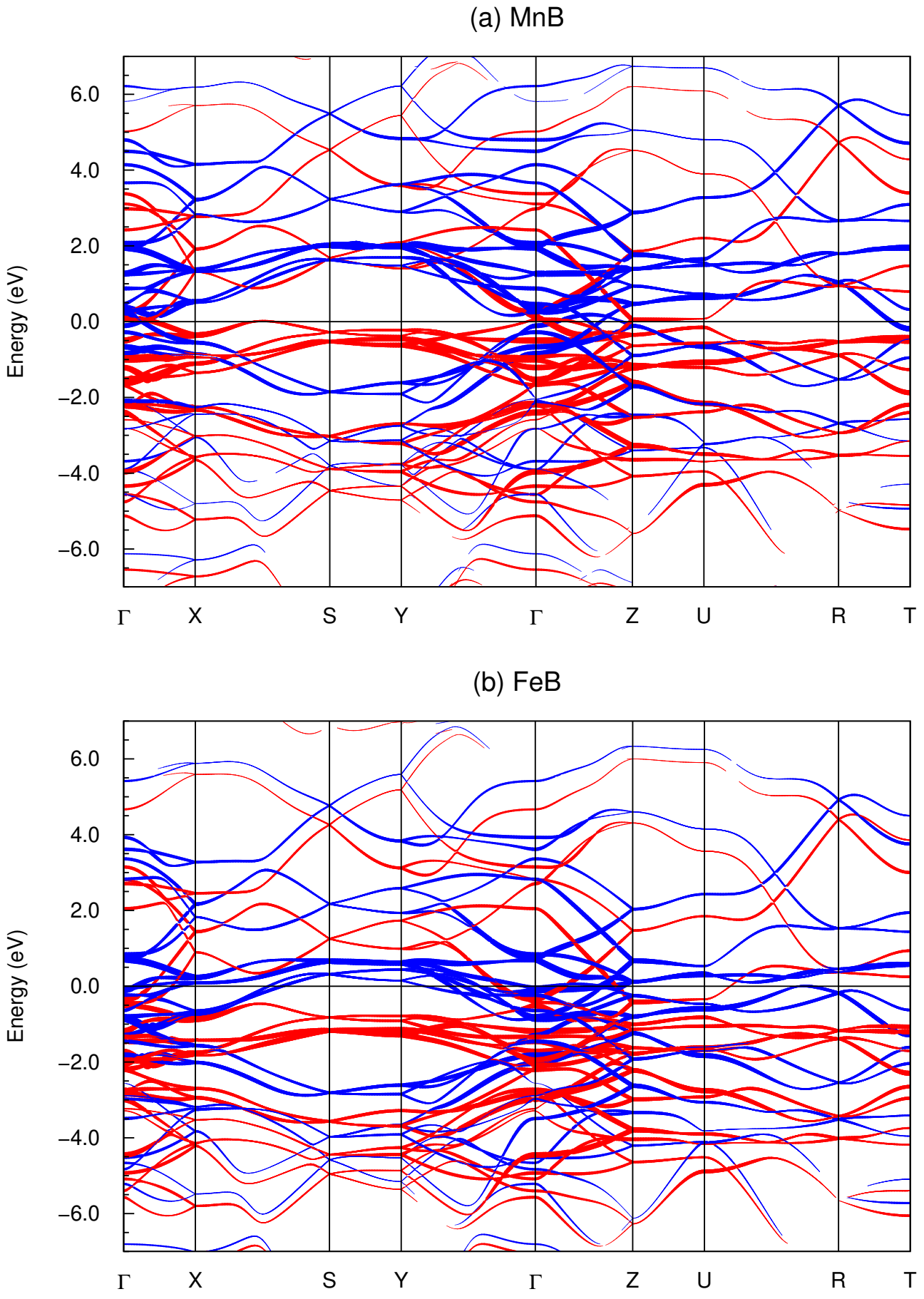

FIG. 4. Fully relativistic $3d$ bands in (a) MnB and (b) FeB. Red and blue bands denote two spin channels. Band thickness represents weighted contribution of $3d$ orbitals. Calculations were performed with the FPLO18 code using the PBE functional.

In Fig. 4, we present weighted contributions from $3d$ orbitals which dominate the band structure of MnB and FeB. We observe a high similarity between the two results. Due to the increase in the number of electrons per formula by one, the Fermi level of FeB rises relative to MnB. Moreover, the spin polarization is noticeably stronger for MnB than for FeB. The change in band filling and spin polarization is also responsible for the evolution of properties in the range of intermediate concentrations between MnB and FeB.

Since the FPLO code used in this work is based on the method of linear combination of atomic orbitals, we can perform a population analysis using the Mulliken approach [76]. In Fig. 5 we show the excess electrons on the

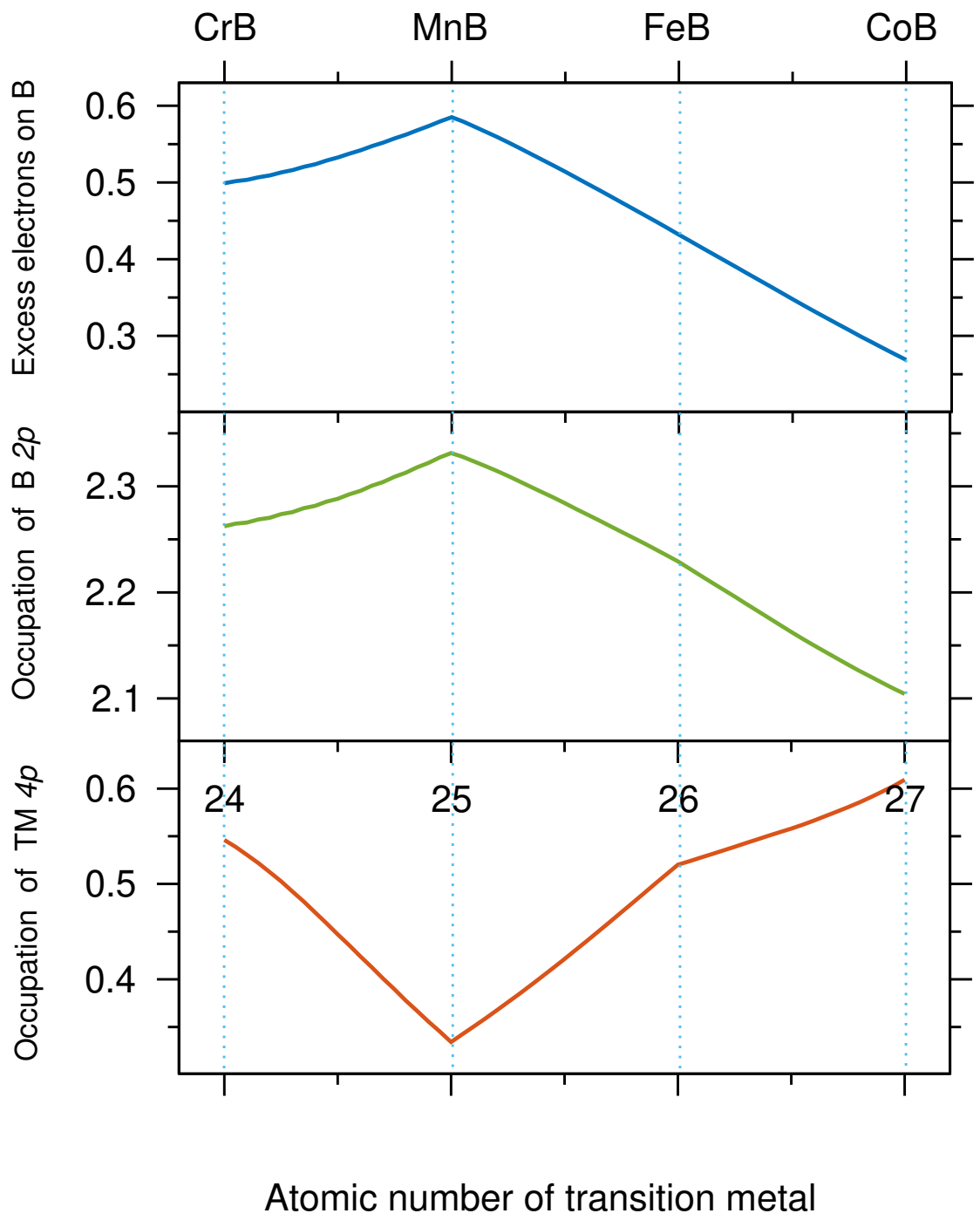

FIG. 5. Excess electrons on boron and occupation of $2p$ orbital of boron and $4p$ orbital of transition metal (TM) as a function of atomic number of transition metal, calculated for the concentration range from CrB, through MnB and FeB, to CoB. The calculations were performed with the virtual crystal approximation (VCA) using the FPLO18 code with PBE exchange-correlation potential.

B site (the natural consequence is the exactly opposite value of excess electrons on the transition metal site). As for the spin magnetic moment dependence, a clear maximum in excess electrons occurs for MnB. Analysis of the occupancy of each orbital shows that the maximum of excess electrons correlates with the occupancy of the B $2p$ and Fe $4p$ orbitals. The occupancy of the other valence orbitals changes linearly or remains constant throughout the whole range of considered alloys (not shown). The densities of states (also not shown) reveal hybridization between the transition metal $4p$ orbitals and boron $2p$ orbitals.

Now we will return to Fig. 2 and analyze the concentration dependence of magnetocrystalline anisotropy energy. The literature on magnetic anisotropy of magnetic monoborides is very scarce. We are only aware of the first-principles study predicting magnetization easy axis for MnB [36] and the experimental work of Zhdanova and coworkers in which they determined the room-temperature effective magnetocrystalline anisotropy constant $K_1^* = 0.4\ \mathrm{MJ\,m^{-3}}$ for orthorhombic FeB [2]. This value leads to a quite promising magnetic hardness of 0.83. The relatively high magnetic hardness is not reflected in the coercivity of FeB, which at 290 K is equal to 10 Oe only [21]. (A similarly low coercivity value, equal to 15.9 Oe, was measured for MnB [10].) However, the relatively high magnetic hardness and low coerciv-ity of FeB do not necessarily contradict each other. As Rades and coworkers showed for $\alpha$-FeB, depending on the size and crystallinity of the particles, soft or hard ferromagnetism is possible to obtain [21]. Therefore, it is conceivable that appropriate control of the microstructure could also improve the coercivity value of the $\beta$-FeB phase.

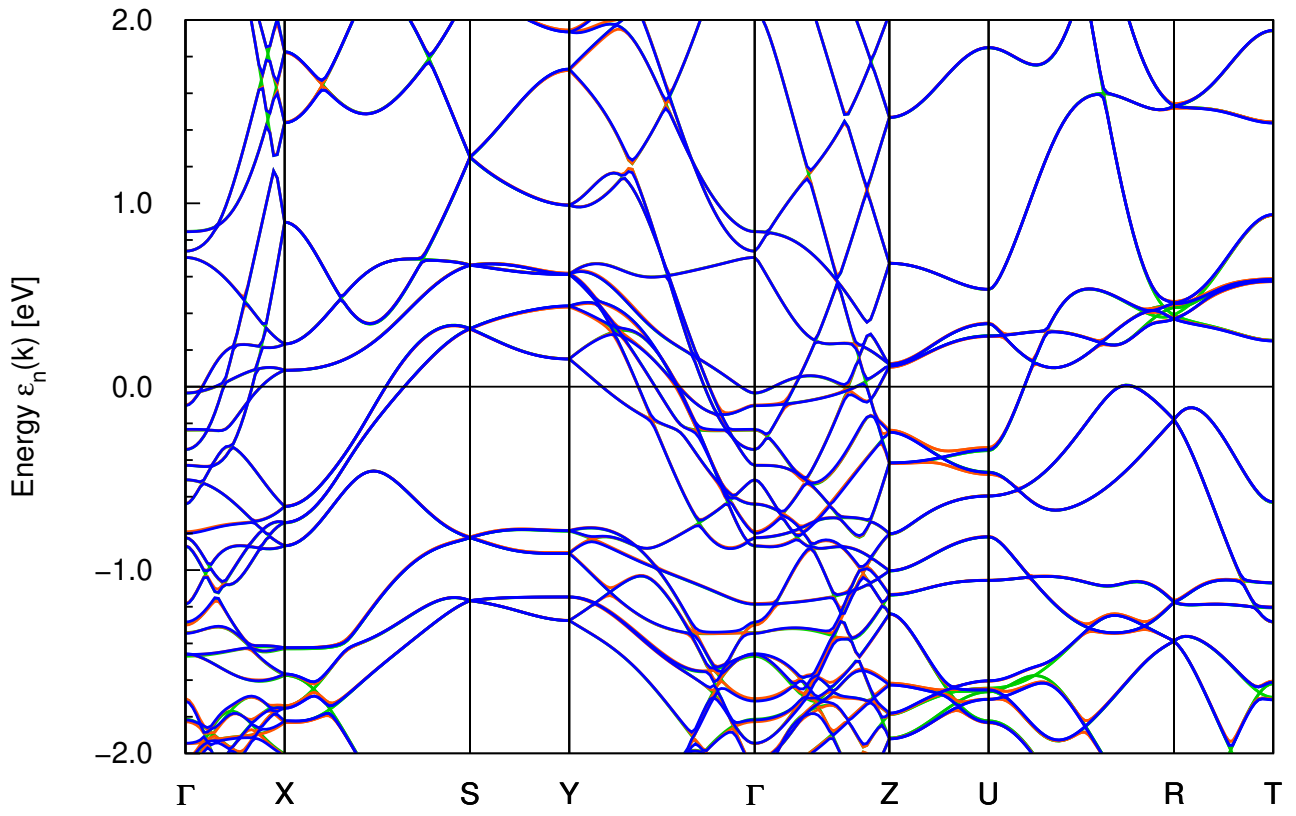

FIG. 6. Band structures calculated for FeB. Red plot for scalar-relativistic solution, green and blue for fully-relativistic solutions with [001] and [010] quantization axes. The calculations were performed with the FPLO18 code with PBE exchange-correlation potential.

In Fig. 2(c), we show the calculated magnetocrystalline anisotropy energy and magnetic hardness. As a result of determining the magnetic easy axis among the three main crystallographic directions, see also Fig. 1, we can divide the analyzed concentration range into areas characterized by a specific easy axis. For MnB and FeB, easy magnetic axis is in both cases [010] (along the axis of lattice parameter $b$) and the values of MAE are of 0.45 and 0.14 $\mathrm{MJ\,m^{-3}}$, respectively, see Table II. In the MnB-FeB range, we observe magnetic hardness maxima at $Mn_{0.55}Fe_{0.45}B$ and $Mn_{0.15}Fe_{0.85}B$ of about 0.83 and 0.95, respectively, with corresponding MAE values equal to 0.77 and 0.69 $\mathrm{MJ\,m^{-3}}$. The highest maximum of MAE equal to 2.9 $\mathrm{MJ\,m^{-3}}$ we observe for $Fe_{0.5}Co_{0.5}B$. Such a high value of anisotropy energy together with a low value of magnetic moment leads to an exceptionally high magnetic hardness of 5.1. Although such a high value of magnetic hardness is determined in an approximate manner from PBE for zero kelvin and probably significantly overestimated by VCA, we still expect experimental confirmation of elevated magnetic hardness in the concentration range from about FeB to $Fe_{0.5}Co_{0.5}B$ at room temperatures. Similarly, very high MAE value (about 10 $\mathrm{MJ\,m^{-3}}$) was predicted previously for rare-earth free FeCo alloys with tetragonal deformation [77].

We determine the MAE as the difference of the fully-relativistic total energies of the system calculated for the magnetic easy axis and the axis perpendicular to it. In Fig. 6, using FeB as an example, we show how small are the differences in band structures for the [001] and [010] quantization axes. In order to make the differences noticeable, we had to narrow the energy window to between -2 and +2 $\mathrm{eV\,atom^{-1}}$. In addition, the figure shows the scalar-relativistic band structure, which is not substantially different from fully-relativistic solutions.

In Fig. 7, we show the band structures calculated for the $Fe_{0.5}Co_{0.5}B$ system with high MAE. Compared to the MnB and FeB band structures presented earlier, we observe further filling of the valence band and a decrease in spin splitting. In a narrow range from -0.6 to 0.6 $\mathrm{eV\,atom^{-1}}$, the differences between the band structures determined for the [001] and [100] axes are visible, as well as the structure of the bands in the vicinity of the Fermi level, which is important from the perspective of the MAE. It is to the Fermi level that the summation of energies for individual $k$-points takes place, see Eq. 2, from which the total MAE can be determined using the magnetic force theorem. The $k$-resolved contributions to the MAE are drawn in green. A lot of positive contributions can be found around the high symmetry point S. It is related to the difference in the [001] and [100] bands splitting observed under the Fermi level near this point. From the point of view of the band structure, we see that in order for the final MAE value to be high, there must be a positive correlation of such factors as spin polarization, the position of the Fermi level, and significant differences in spin-orbit splitting for different quantization axes.

We must keep in mind that the magnetocrystalline anisotropy energies presented in this work were calculated in zero kelvin and may change with temperature. As previous measurements for halfborides have shown, with temperature the anisotropy constants can change in a nonlinear fashion [4, 78]. Nevertheless, values close to the anisotropy energy maximum observed for low temperatures in the concentration range reappear in the temperature dependence of the anisotropy constants. We conclude, that also the maxima in the measured temperature dependence of the anisotropy constants for monoborides should have values close to the anisotropy energy maxima calculated at zero kelvin. For the MnB-FeB-CoB alloys considered here, it would be advantageous to experimentally verify the calculation results extended by measuring the temperature dependence of the anisotropy constants.

### B. Fixed spin moment on FeB

In our previous work on $CeFe_{12}$-based alloys, we showed that the calculated MAE value depends on the choice of the exchange-correlation potential [67], and that this dependency is correlated with the magnetic moment. Here, we performed MAE calculations for the selected FeB compound using four different forms of exchange-correlation potential: Perdew-Burke-Ernzerhof (PBE), Barth-Hedin (BH), Perdew-Wang (PW92), and *exchange only* LDA (for comparison only). The obtained equilibrium results, see Fig. 8 (green dots) and Table III, show the scatter of the magnetic moments and MAE. The total spin magnetic moment from GGA-PBE is 1.20 $\mu_B$ $\mathrm{f.u.^{-1}}$, whereas the experimental one is 1.15 $\mu_B$ f.u.$^{-1}$ [1]. The induced spin magnetic moments on B are proportional to the moments on Fe, see Table III. The total spin magnetic moments from the GGA and proper (not exchange only) LDA differ by about 0.1 $\mu_B$, which correlates with a MAE variance of about 0.13 MJ m$^{-3}$. However, even the higher MAE value of 0.14 MJ m$^{-3}$ determined from the PBE approximation is much smaller than the experimental effective magnetic anisotropy constant $K_1^* = 0.4$ MJ m$^{-3}$ [2]. In addition to the approximation of exchange-correlation potential, the divergence may have its origin in slightly different definitions of MAE and $K_1^*$, as well as in the temperature dependence of the parameters. We calculated MAE at 0 K and in the experiment $K_1^*$ is determined at room temperature.

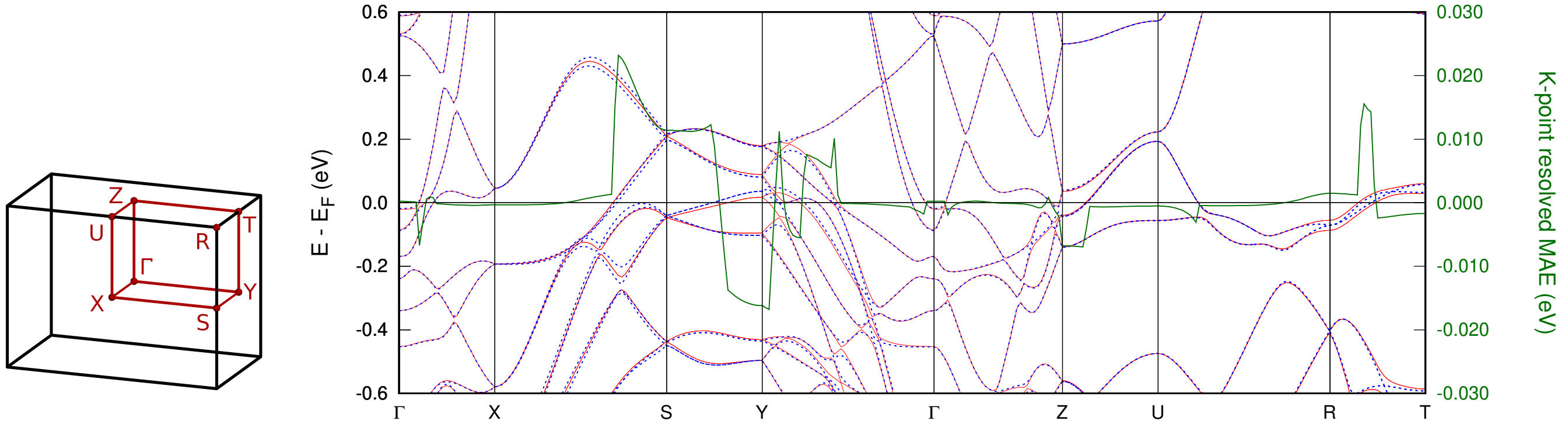

FIG. 7. Band structure for $Fe_{0.5}Co_{0.5}B$ – an alloy with high magnetocrystalline anisotropy energy (MAE) equal to 2.87 MJ m$^{-3}$. Dashed blue lines indicate solution for [001] quantization axis, whereas solid red lines stand for solution for [100] axis. Green lines denote $k$-point-resolved contributions to MAE. Calculations were performed with the FPLO18 code using the PBE functional and virtual crystal approximation (VCA). Brillouin zone with high symmetry points is shown.

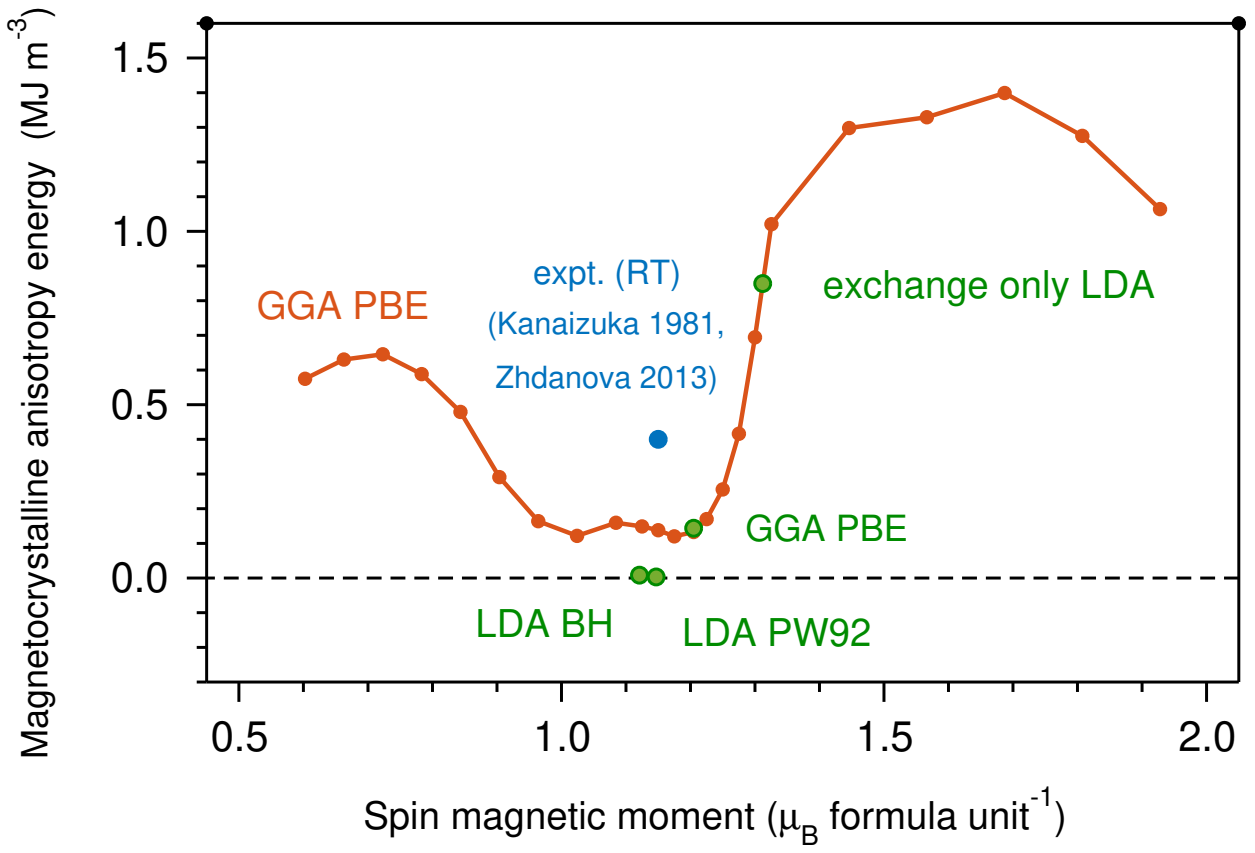

FIG. 8. The dependence of the magnetocrystalline anisotropy energy (MAE) on a fixed spin moment calculated for FeB (s.g. *Pnma*) using the FPLO18 code with Perdew-Burke-Ernzerhof (PBE) exchange-correlation potential; together with the points of equilibrium values of MAE and equilibrium values of spin magnetic moments obtained with GGA PBE, LDA Barth-Hedin (LDA BH), LDA Perdew-Wang (LDA PW92), and LDA *exchange only* potentials. The experimental (room-temperature) values of magnetic moment and effective magnetic anisotropy constant are taken from Refs. [1] and [2].

TABLE III. Magnetic properties of FeB (*Pnma*) calculated with FPLO18 code using several exchange-correlation potentials: Perdew-Burke-Ernzerhof (PBE), Barth-Hedin (BH), Perdew-Wang (PW92), and LDA *exchange only*. The magnetic easy axis, spin ($m_s$) and orbital ($m_l$) magnetic moments ($\mu_B$ atom$^{-1}$), magnetocrystalline anisotropy energy [MAE (MJ m$^{-3}$)], and magnetic hardness ($\kappa$) are presented. Orbital magnetic moments are given for corresponding magnetic easy axes.

| XC potential | Axis | $m_s$(Fe) | $m_l$(Fe) | $m_s$(B) | $m_l$(B) | MAE | $\kappa$ |
|---|---|---|---|---|---|---|---|
| GGA PBE | 010 | 1.33 | 0.025 | -0.13 | 0.001 | 0.14 | 0.49 |
| LDA BH | 100 | 1.21 | 0.025 | -0.09 | 0.000 | 0.01 | 0.13 |
| LDA PW92 | 010 | 1.25 | 0.023 | -0.10 | 0.000 | 0.01 | 0.08 |
| LDA ex. only | 010 | 1.46 | 0.024 | -0.15 | 0.001 | 0.85 | 1.10 |

As we can see from the above calculations, there may be a correlation between the determined MAE value and the magnetic moment. To verify this, we performed a fully-relativistic fixed spin magnetic moment (FSM) calculation to determine the dependence of the MAE on the magnetic moment. In Fig. 8, we can see a fairly good correspondence between the equilibrium points calculated using different forms of exchange-correlation potential (green dots) and the curve of the MAE dependence on fixed spin magnetic moment (red line). We also see that the equilibrium results, except LDA *exchange only*, lie between the two MAE maxima, and both increasing and decreasing the magnetic moment of the system can lead to an increase in the MAE [characteristic also visible around FeB in Fig. 2(c)]. In practice, the magnetic moment of FeB can be decreased by alloying with Co and increased by alloying with Mn (or small amounts of Cr) [6]. The value of the magnetic moment is also affected by the temperature, hence the dependence of MAE on temperature can be nonlinear [3, 4].

To better understand the evolution of MAE with the fixed spin moment, we take a closer look at cases with fixed spin moments ($m_s$) equal to 0.72, 1.21, and 1.57 $\mu_B$ f.u.$^{-1}$, of which the first and last are characterized by high MAE values, and the middle one is an equilibrium solution with a low MAE, see Fig. 9. The densities of states (DOS) span a relatively wide energy range from -7 to 7 eV atom$^{-1}$, covering most of the Fe 3$d$ band. As the magnetic moment increases, the majority (upper) spin channel shifts to the left, and the minority band remains nearly stationary. Therefore, we associate the observed evolution of the MAE with a fixed spin moment with the shifting of the Fermi level along the majority band. In the equilibrium state, see Fig. 9(b), the Fermi level of the majority spin channel falls between the non-bonding and anti-bonding states, as Mohn and Pettifor have noted [38]. In contrast, in the two cases with high MAE, the Fermi level is at the non-bonding and anti-bonding peak. As the presented spin-polarized DOS were resolved by orbital magnetic quantum number $m$, in the case with the highest MAE ($m_s = 1.57$ $\mu_B$ f.u.$^{-1}$) we see that the most pronounced contribution to the majority band on the Fermi level comes from the states with quantum number $m = 2$.

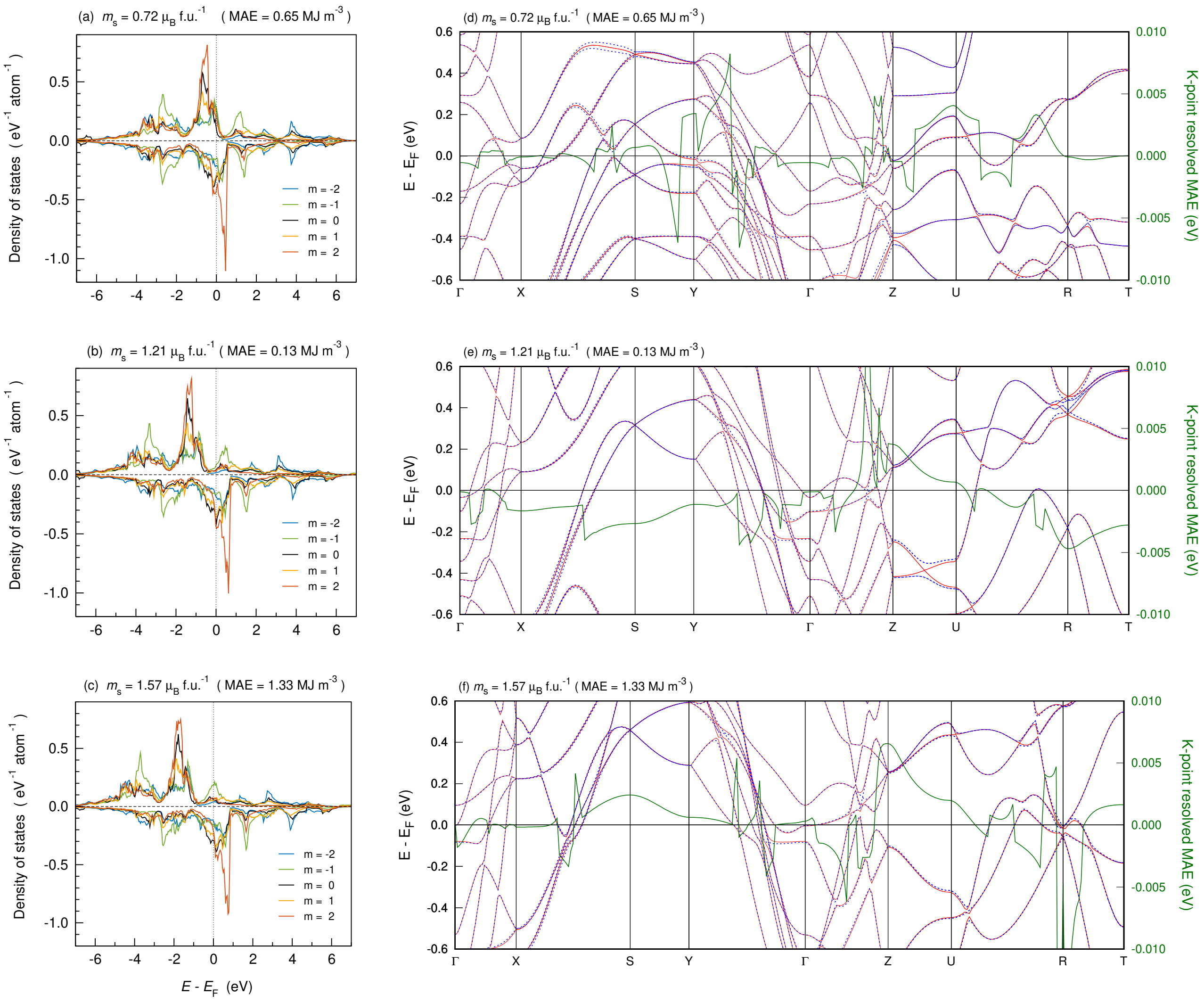

FIG. 9. Densities of states, band structures, and contribution to magnetocrystalline anisotropy energy (MAE) calculated for fixed spin moments of 0.72, 1.21, and 1.57 $\mu_B$ f.u.$^{-1}$ for FeB (s.g. *Pnma*). Spin-polarized densities of Fe 3$d$ states were resolved by orbital magnetic quantum number $m$. Dashed blue lines represent band structures calculated for [010] quantization axis, whereas solid red lines stand for results obtained for [100] axis (in case of configurations with $m_s$ equal to 0.72 and 1.21 $\mu_B$ f.u.$^{-1}$) and [001] axis (in case of system with $m_s = 1.57$ $\mu_B$ f.u.$^{-1}$). Green lines denote $k$-point-resolved contributions to MAE. Calculations were performed with the FPLO18 code using the PBE functional.

The second time we see the Fermi level scanning the majority band is in the band structure plots, see Fig. 9(d-f). However, this time in the much narrower energy range of -0.6 to 0.6 eV atom$^{-1}$. With this form of presentation and the determination of contributions to the MAE from individual $k$-points, we are additionally able to identify high-symmetry points around which large positive areas of MAE have formed. For $m_s = 0.72$ $\mu_B$ f.u.$^{-1}$, these are mainly Y and U points, for $m_s = 1.21$ $\mu_B$ f.u.$^{-1}$ point Z, and for $m_s = 1.57$ $\mu_B$ f.u.$^{-1}$ points S and Z.

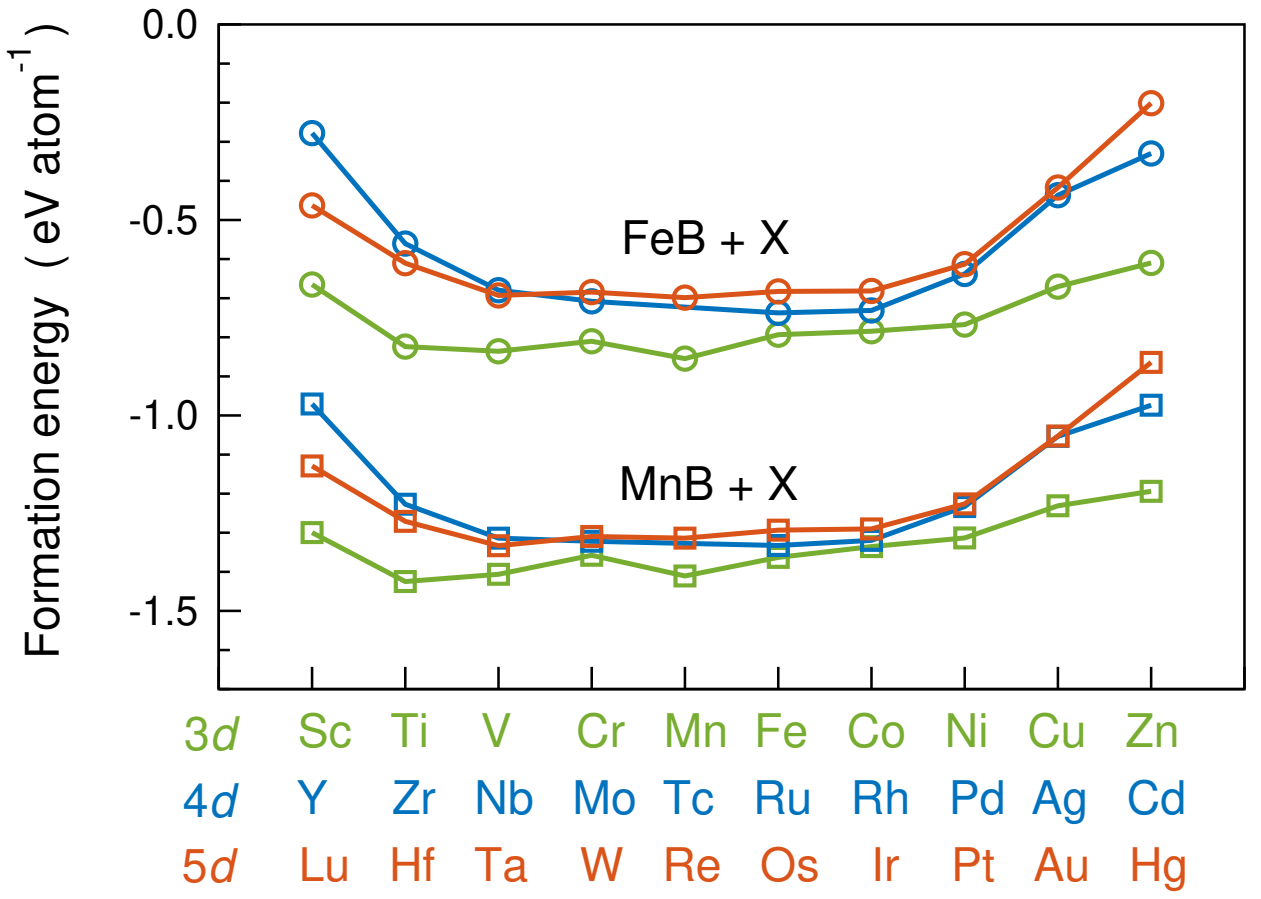

FIG. 10. Circles and squares indicate formation energies for $Fe_{11}X_1B_{12}$ and $Mn_{11}X_1B_{12}$ alloys, respectively, with 3*d*, 4*d*, and 5*d* transition metals (X). The calculations were performed with FPLO18 code using the PBE functional and supercell approach.

This type of analysis of the dependence of MAE on magnetic moment also provides a clue on how to interpret MAE changes with temperature. This is because with temperature, the magnetic moment of the system changes, which can lead to a similar evolution of MAE as discussed above.

The analysis of the PBE results against the experiment and other exchange-correlation functionals lends credence to the use of the PBE approximation, while pointing out its inherent limitations. Similar to the VCA findings presented above, the fixed spin moment results are promising regarding application of monoborides as permanent magnets and suggest the possibility of modifying the composition to increase the magnetic hardness of the alloys above unity.

Although it is often difficult to control the details of the band structure to achieve optimal material properties, the regularities observed above offer hope for the development of MAE engineering based on such control.

### C. MnB and FeB alloys with 3*d*, 4*d*, and 5*d* transition metals

We used the supercell method to calculate the properties of series of $Mn_{11}X_1B_{12}$ and $Fe_{11}X_1B_{12}$ alloys with transition metals X. We assumed that the transition metal atom substitutes one of the twelve Mn or Fe atoms in the supercell, see also Sec. II: *Computational details*.

Before we move on to the magnetic properties of these systems, we will first examine their chemical stability. As we see in Fig. 10, all the systems under consideration have negative formation energy, see Eq. 3, which means they are all chemically stable. Manganese-based compositions have lower energies than Fe-based compositions, and transition-metal substitutions from the edges of the periods noticeably increase the formation energy (lowering stability). The calculated formation energy values, in the range down to approximately -1.5 eV atom$^{-1}$, indicate a good potential of the monoborides to be substituted by most of the transition metal elements.

A previous DFT study of the alloying of $(Fe,Co)_2B$ halfborides with 5*d* elements predicted a significant increase in MAE for Re doping (from about 0.75 to 1.4 MJ m$^{-3}$), which found partial experimental confirmation (increase from 0.5 to 0.75 MJ m$^{-3}$) [3]. Here, we will discuss how the substitution of MnB and FeB with transition metals affects the magnetic hardness of the monoborides. The results of calculated magnetic moments and magnetocrystalline anisotropy energies are presented in Table IV and Figs. 11 and 12. In most cases, we see clear trends in the dependence on the atomic number of the transition metal. The trends are often analogous to the VCA results presented earlier in this work and supercell results from our previous work on $CeFe_{12}$ based alloys [67].

Figure 11 presents the calculated magnetic properties of MnB alloys. Figure 11(a) shows that of all the systems, the highest total spin magnetic moment has the MnB without substitutions. It is analogous to the experimental results discussed above [1, 6]. We can also see that the magnetic moment dependencies on the type of substitution are similar for all three transition metal groups: 3*d*, 4*d* and 5*d*. This is because, when we substitute Mn with transition metals X we change the filling of the valence band.

Figure 11(b) shows the spin magnetic moments at the dopant site. We can see that transition metal atoms are not solely non-magnetic dopants that lower the total magnetic moment, but they undergo spin polarization in the ferromagnetic medium and contribute to the total magnetic moment. The magnetic contributions are positive or negative (parallel or antiparallel to the total magnetic moment), with a maxima near the center of the periods. The considered dupants also affect the magnetic interactions between Mn atoms.

Figure 11(c) shows the orbital magnetic moments on the dopant site. The values of these moments oscillate between around -0.03 and +0.05 $\mu_B$ atom$^{-1}$, which is small compared to the spin magnetic moments of about 1.6 – 2.0 $\mu_B$ f.u.$^{-1}$ in the MnB alloys.

Figures 11(d) and 11(f) show the magnetocrystalline anisotropy energies (MAE and $DE_{32}$). As previously mentioned, the direction of the magnetic easy axis for MnB is [010]. In most cases, substituting MnB with transition metals changes this preference to [001], see Table IV. MnB has the highest MAE value of 0.67 MJ m$^{-3}$. This value translates into magnetic hardness of 0.7, see Fig. 11(e), and no dopant raises the magnetic hardness above 0.8. Based on the above, we conclude that all Mn monoborides doped with transition metals ($Mn_{11}X_1B_{12}$) can be classified as soft ($\kappa < 0.5$) or semi-hard ($0.5 < \kappa < 1$) magnetic materials.

The observed trends of magnetic moments for $Fe_{11}X_1B_{12}$ alloys, see Fig. 12 and Table IV, resemble those just presented for MnB alloys. However, the mag-

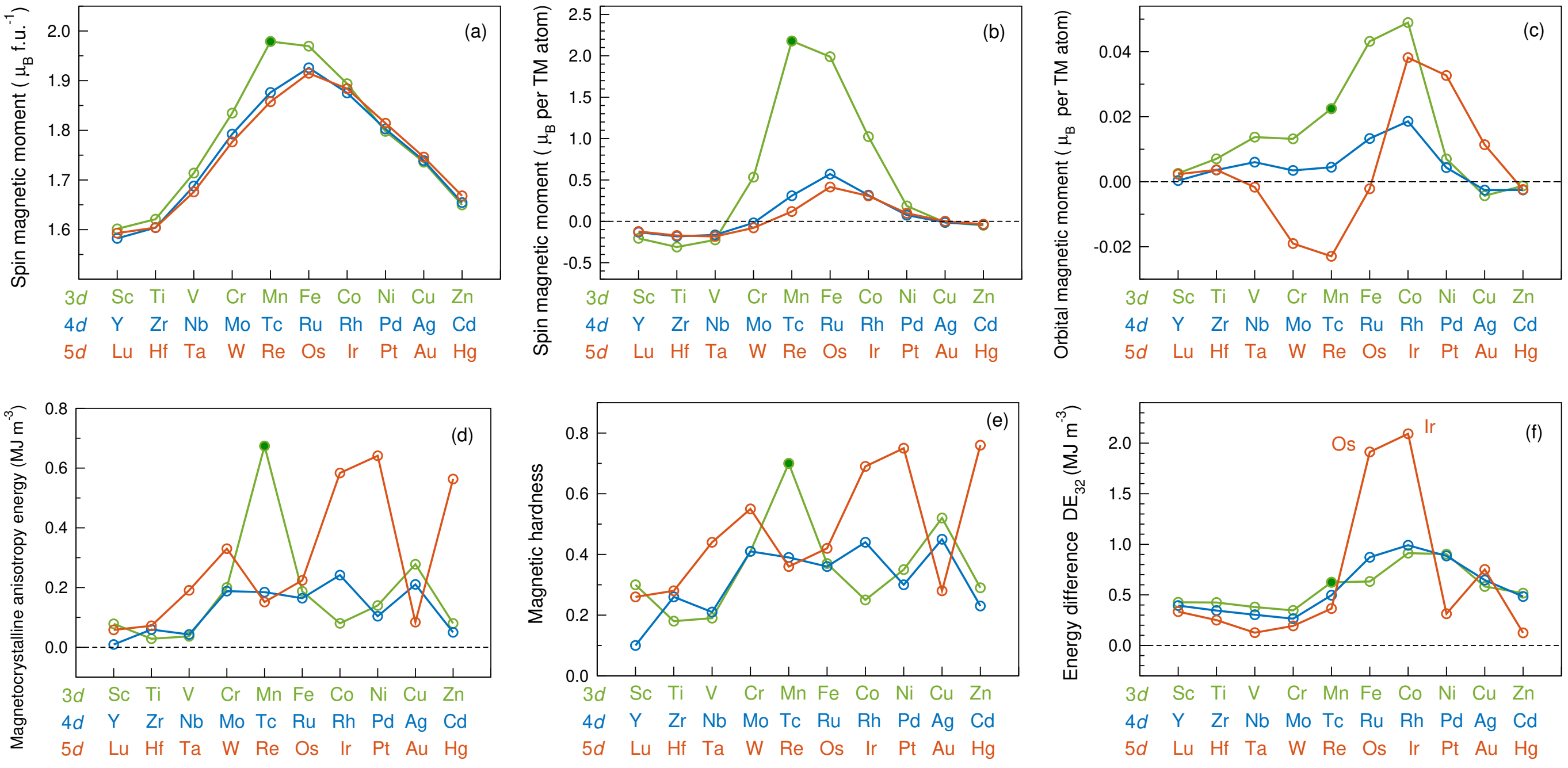

FIG. 11. The magnetic properties of MnB alloys with 3$d$, 4$d$, and 5$d$ transition metals X ($Mn_{11}X_1B_{12}$): (a) total spin magnetic moment (per formula unit of two atoms); (b, c) spin and orbital magnetic moments on dopant atoms X (per TM atom); (d) magnetocrystalline anisotropy energy (MAE); (e) magnetic hardness; and (f) energy difference ($DE_{32}$) between the two higher energies among the three total energies calculated along the three main crystallographic axes. The calculations were performed with FPLO18 code using the PBE functional and supercell approach. Full circles indicate results for the base material (MnB) without dopant of other type.

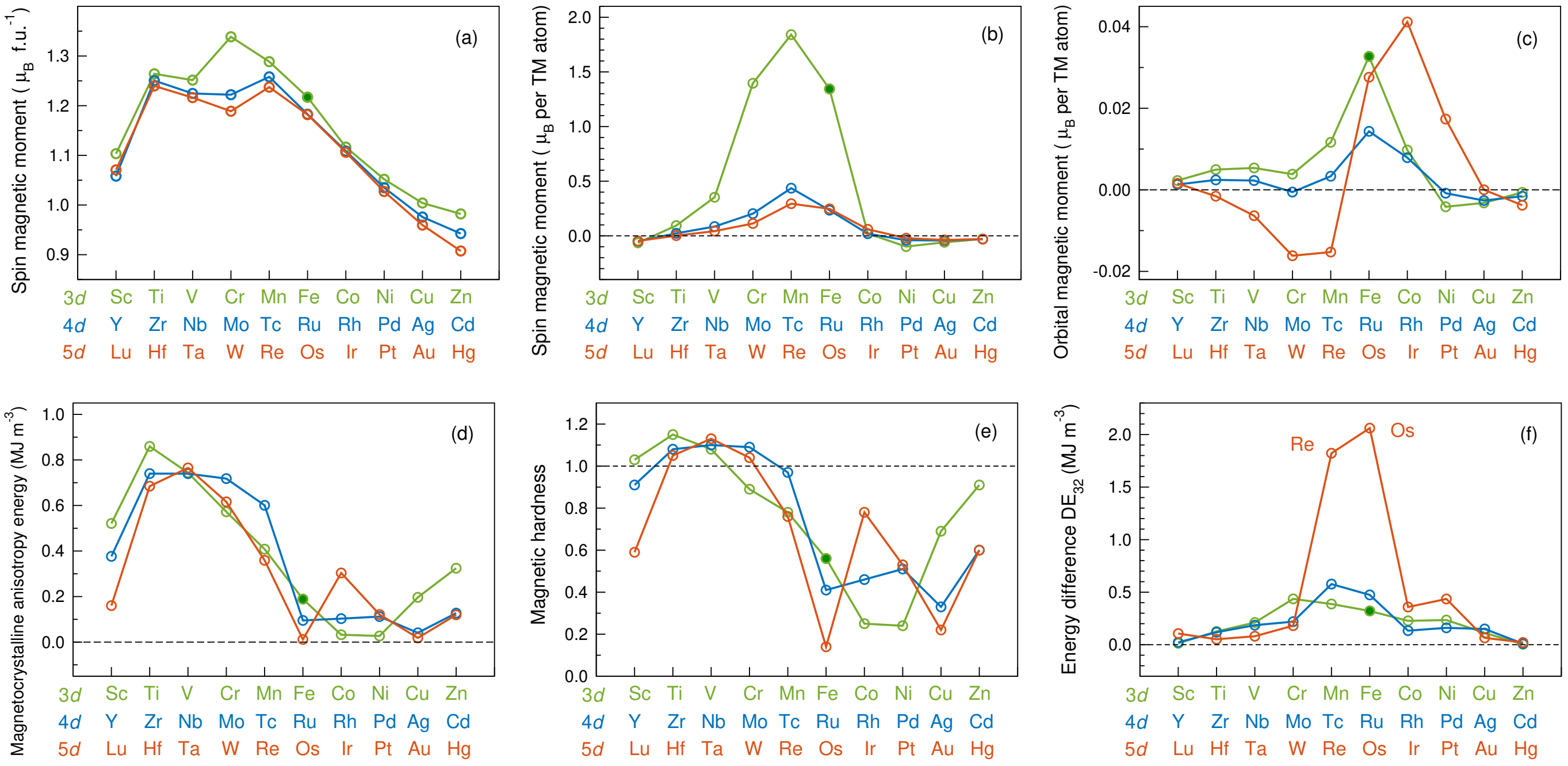

FIG. 12. The magnetic properties of FeB alloys with 3$d$, 4$d$, and 5$d$ transition metals X ($Fe_{11}X_1B_{12}$): (a) total spin magnetic moment (per formula unit of two atoms); (b, c) spin and orbital magnetic moments on dopant atoms X (per TM atom); (d) magnetocrystalline anisotropy energy (MAE); (e) magnetic hardness; and (f) energy difference ($DE_{32}$) between the two higher energies among the three total energies calculated along the three main crystallographic axes. The calculations were performed with FPLO18 code using the PBE functional and supercell approach. Full circles indicate results for the base material (FeB) without dopant of other type.

TABLE IV. The magnetic easy axis, magnetocrystalline anisotropy energy [MAE (MJ m$^{-3}$)], total spin magnetic moment [$m$ ($\mu_B$ f.u.$^{-1}$); formula unit consists of two atoms], magnetic hardness [$\kappa$], orbital magnetic moment [$m_l$ ($\mu_B$ per TM atom)] for [001] quantization axes, and spin magnetic moment on substituting atom [$m_s$(X) ($\mu_B$ per TM atom)] calculated for $Mn_{11}X_1B_{12}$ and $Fe_{11}X_1B_{12}$ supercells with various $3d$, $4d$, and $5d$ transition metal elements X. Calculations were performed with the FPLO18 code using the PBE functional.

| $Mn_{11}XB_{12}$ | | | | | | | $Fe_{11}XB_{12}$ | | | | | | |
|---|---|---|---|---|---|---|---|---|---|---|---|---|---|
| | | | | | | | $3d$ elements | | | | | | |
| Alloy | Axis | MAE | $m$ | $\kappa$ | $m_l(X)$ | $m_s(X)$ | Alloy | Axis | MAE | $m$ | $\kappa$ | $m_l(X)$ | $m_s(X)$ |
| $Mn_{0.917}Sc_{0.083}B$ | 001 | 0.08 | 1.60 | 0.30 | 0.003 | -0.21 | $Fe_{0.917}Sc_{0.083}B$ | 010 | 0.52 | 1.10 | 1.03 | 0.002 | -0.06 |
| $Mn_{0.917}Ti_{0.083}B$ | 010 | 0.03 | 1.62 | 0.18 | 0.007 | -0.31 | $Fe_{0.917}Ti_{0.083}B$ | 010 | 0.86 | 1.26 | 1.15 | 0.005 | 0.09 |
| $Mn_{0.917}V_{0.083}B$ | 001 | 0.04 | 1.71 | 0.19 | 0.014 | -0.22 | $Fe_{0.917}V_{0.083}B$ | 010 | 0.74 | 1.25 | 1.08 | 0.005 | 0.35 |
| $Mn_{0.917}Cr_{0.083}B$ | 001 | 0.20 | 1.83 | 0.41 | 0.013 | 0.53 | $Fe_{0.917}Cr_{0.083}B$ | 010 | 0.57 | 1.34 | 0.89 | 0.004 | 1.40 |
| MnB | 010 | 0.67 | 1.98 | 0.70 | 0.022 | 2.18 | $Fe_{0.917}Mn_{0.083}B$ | 010 | 0.41 | 1.29 | 0.78 | 0.012 | 1.84 |
| $Mn_{0.917}Fe_{0.083}B$ | 001 | 0.19 | 1.97 | 0.78 | 0.043 | 1.99 | FeB | 010 | 0.19 | 1.22 | 0.56 | 0.033 | 1.34 |
| $Mn_{0.917}Co_{0.083}B$ | 001 | 0.08 | 1.89 | 0.25 | 0.049 | 1.02 | $Fe_{0.917}Co_{0.083}B$ | 100 | 0.03 | 1.12 | 0.25 | 0.010 | 0.02 |
| $Mn_{0.917}Ni_{0.083}B$ | 001 | 0.14 | 1.80 | 0.35 | 0.007 | 0.19 | $Fe_{0.917}Ni_{0.083}B$ | 010 | 0.03 | 1.05 | 0.24 | -0.004 | -0.10 |
| $Mn_{0.917}Cu_{0.083}B$ | 001 | 0.28 | 1.74 | 0.52 | -0.004 | -0.01 | $Fe_{0.917}Cu_{0.083}B$ | 010 | 0.2 | 1.00 | 0.69 | -0.003 | -0.06 |
| $Mn_{0.917}Zn_{0.083}B$ | 001 | 0.08 | 1.65 | 0.29 | -0.001 | -0.05 | $Fe_{0.917}Zn_{0.083}B$ | 010 | 0.32 | 0.98 | 0.91 | -0.001 | -0.03 |
| | | | | | | | $4d$ elements | | | | | | |
| $Mn_{0.917}Y_{0.083}B$ | 010 | 0.01 | 1.58 | 0.10 | 0.000 | -0.13 | $Fe_{0.917}Y_{0.083}B$ | 010 | 0.38 | 1.06 | 0.91 | 0.001 | -0.05 |
| $Mn_{0.917}Zr_{0.083}B$ | 010 | 0.06 | 1.60 | 0.26 | 0.004 | -0.18 | $Fe_{0.917}Zr_{0.083}B$ | 010 | 0.74 | 1.25 | 1.08 | 0.002 | 0.02 |
| $Mn_{0.917}Nb_{0.083}B$ | 001 | 0.04 | 1.69 | 0.21 | 0.006 | -0.16 | $Fe_{0.917}Nb_{0.083}B$ | 010 | 0.74 | 1.22 | 1.10 | 0.002 | 0.08 |
| $Mn_{0.917}Mo_{0.083}B$ | 001 | 0.19 | 1.79 | 0.41 | 0.003 | -0.02 | $Fe_{0.917}Mo_{0.083}B$ | 010 | 0.72 | 1.22 | 1.09 | -0.001 | 0.20 |
| $Mn_{0.917}Tc_{0.083}B$ | 001 | 0.18 | 1.88 | 0.39 | 0.004 | 0.31 | $Fe_{0.917}Tc_{0.083}B$ | 010 | 0.60 | 1.26 | 0.97 | 0.003 | 0.44 |
| $Mn_{0.917}Ru_{0.083}B$ | 001 | 0.16 | 1.93 | 0.36 | 0.013 | 0.57 | $Fe_{0.917}Ru_{0.083}B$ | 010 | 0.10 | 1.18 | 0.41 | 0.014 | 0.23 |
| $Mn_{0.917}Rh_{0.083}B$ | 001 | 0.24 | 1.88 | 0.44 | 0.019 | 0.32 | $Fe_{0.917}Rh_{0.083}B$ | 100 | 0.10 | 1.11 | 0.46 | 0.008 | 0.02 |
| $Mn_{0.917}Pd_{0.083}B$ | 001 | 0.10 | 1.80 | 0.30 | 0.004 | 0.07 | $Fe_{0.917}Pd_{0.083}B$ | 100 | 0.11 | 1.04 | 0.51 | -0.001 | -0.04 |
| $Mn_{0.917}Ag_{0.083}B$ | 001 | 0.21 | 1.74 | 0.45 | -0.003 | -0.01 | $Fe_{0.917}Ag_{0.083}B$ | 010 | 0.04 | 0.98 | 0.33 | -0.003 | -0.04 |
| $Mn_{0.917}Cd_{0.083}B$ | 010 | 0.05 | 1.65 | 0.23 | -0.003 | -0.04 | $Fe_{0.917}Cd_{0.083}B$ | 010 | 0.13 | 0.94 | 0.60 | -0.002 | -0.03 |
| | | | | | | | $5d$ elements | | | | | | |
| $Mn_{0.917}Lu_{0.083}B$ | 001 | 0.06 | 1.59 | 0.26 | 0.002 | -0.12 | $Fe_{0.917}Lu_{0.083}B$ | 010 | 0.16 | 1.07 | 0.59 | 0.002 | -0.05 |
| $Mn_{0.917}Hf_{0.083}B$ | 010 | 0.07 | 1.60 | 0.28 | 0.004 | -0.17 | $Fe_{0.917}Hf_{0.083}B$ | 010 | 0.68 | 1.24 | 1.05 | -0.002 | 0.00 |
| $Mn_{0.917}Ta_{0.083}B$ | 001 | 0.19 | 1.68 | 0.44 | -0.002 | -0.18 | $Fe_{0.917}Ta_{0.083}B$ | 010 | 0.77 | 1.22 | 1.13 | -0.006 | 0.04 |
| $Mn_{0.917}W_{0.083}B$ | 001 | 0.33 | 1.78 | 0.55 | -0.019 | -0.08 | $Fe_{0.917}W_{0.083}B$ | 010 | 0.62 | 1.19 | 1.04 | -0.016 | 0.11 |
| $Mn_{0.917}Re_{0.083}B$ | 010 | 0.15 | 1.86 | 0.36 | -0.023 | 0.12 | $Fe_{0.917}Re_{0.083}B$ | 010 | 0.36 | 1.24 | 0.76 | -0.015 | 0.29 |
| $Mn_{0.917}Os_{0.083}B$ | 010 | 0.22 | 1.91 | 0.42 | -0.002 | 0.41 | $Fe_{0.917}Os_{0.083}B$ | 010 | 0.01 | 1.18 | 0.14 | 0.028 | 0.25 |
| $Mn_{0.917}Ir_{0.083}B$ | 001 | 0.58 | 1.88 | 0.69 | 0.038 | 0.31 | $Fe_{0.917}Ir_{0.083}B$ | 100 | 0.30 | 1.11 | 0.78 | 0.041 | 0.06 |
| $Mn_{0.917}Pt_{0.083}B$ | 001 | 0.64 | 1.81 | 0.75 | 0.033 | 0.10 | $Fe_{0.917}Pt_{0.083}B$ | 100 | 0.12 | 1.03 | 0.53 | 0.017 | -0.02 |
| $Mn_{0.917}Au_{0.083}B$ | 001 | 0.08 | 1.75 | 0.28 | 0.011 | 0.00 | $Fe_{0.917}Au_{0.083}B$ | 100 | 0.02 | 0.96 | 0.22 | 0.000 | -0.04 |
| $Mn_{0.917}Hg_{0.083}B$ | 010 | 0.56 | 1.67 | 0.76 | -0.002 | -0.03 | $Fe_{0.917}Hg_{0.083}B$ | 010 | 0.12 | 0.91 | 0.60 | -0.004 | -0.03 |

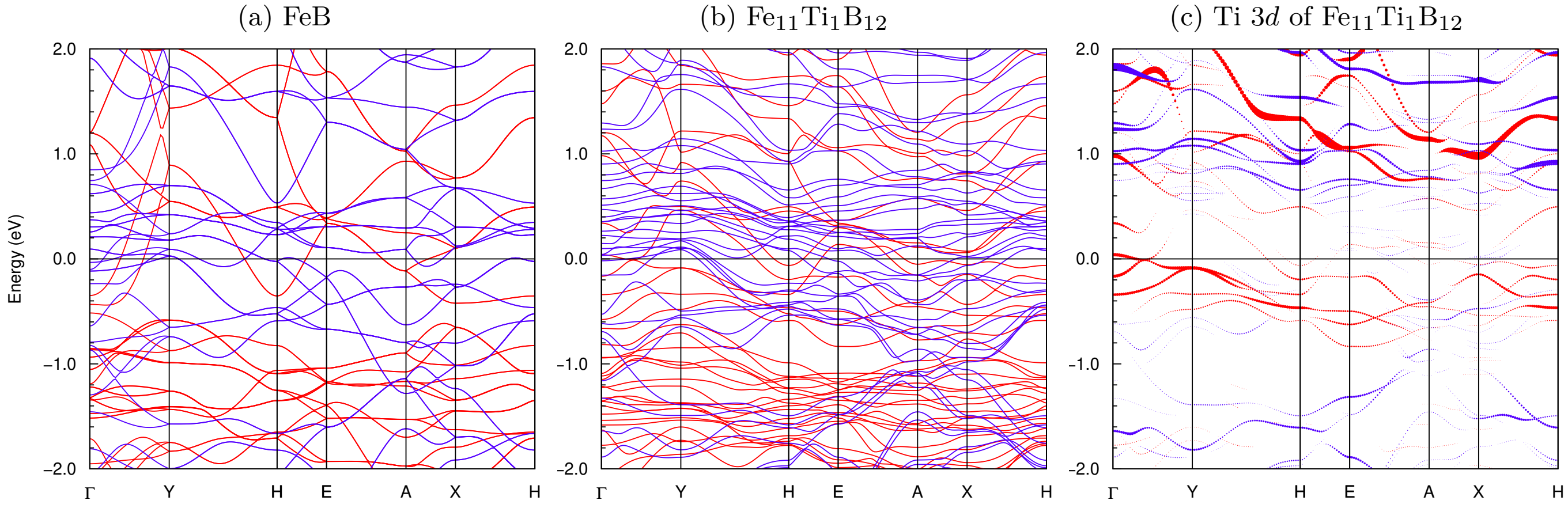

FIG. 13. Band structure of (a) FeB, (b,c) $Fe_{11}Ti_1B_{12}$ and weighted contributions of Ti $3d$ orbitals. Red and blue color denote two spin channels. The calculations were performed with FPLO18 code using the PBE functional and supercell approach.

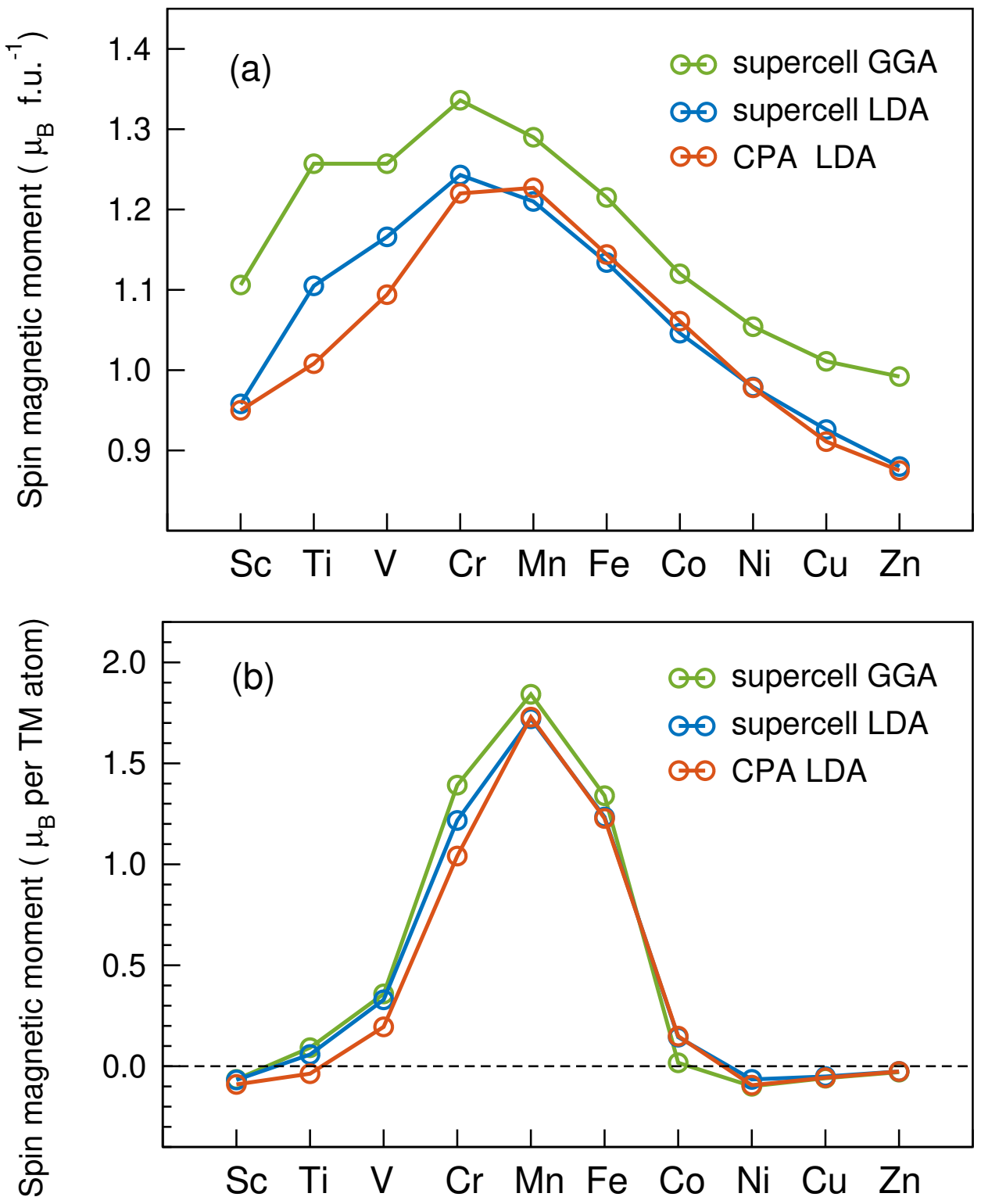

FIG. 14. The magnetic moments of FeB alloys with $3d$ transition metals X ($Fe_{11}X_1B_{12}$): (a) total spin magnetic moment (per formula unit of two atoms) and (b) spin magnetic moments on dopant atoms X (per TM atom). The supercell calculations were performed with FPLO18 code using the GGA-PBE and LDA-PW92 functionals. The calculations with coherent potential approximation (CPA) were carried out with FPLO5 and LDA-PW92 functional.

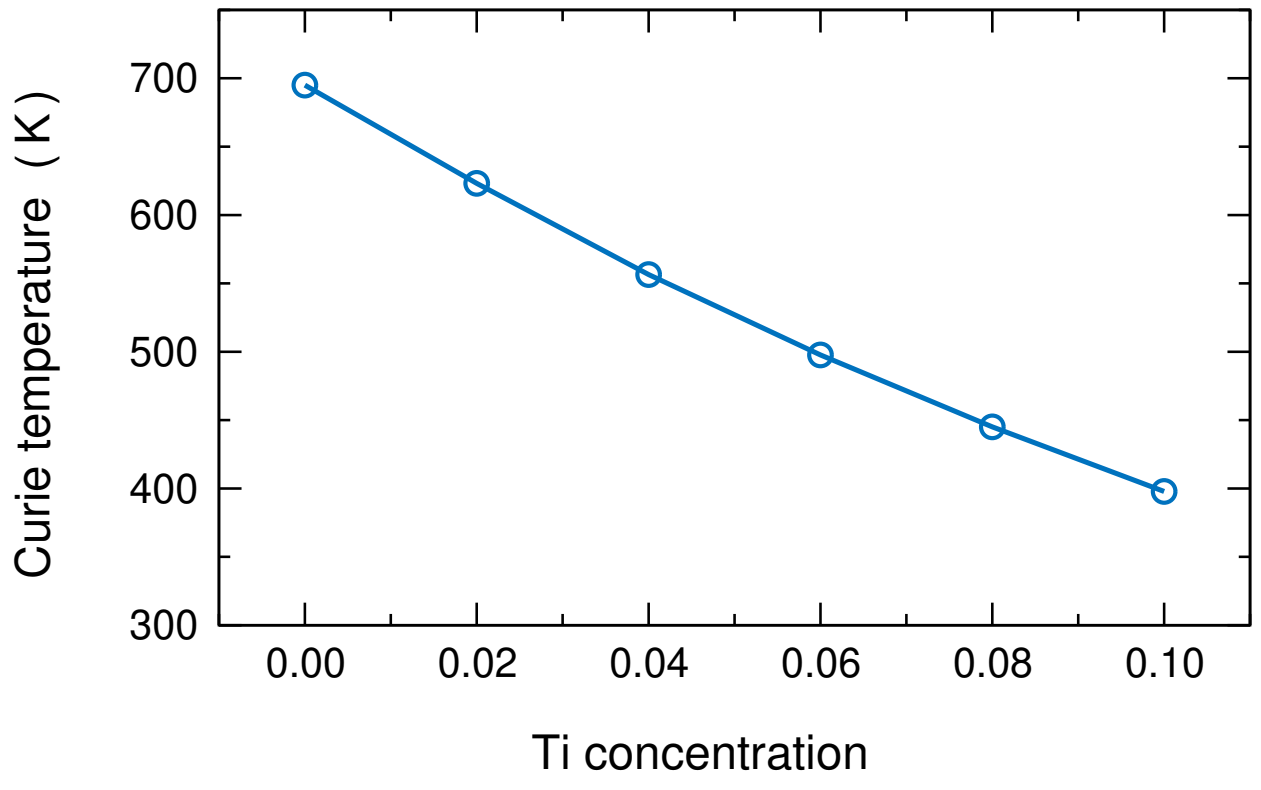

FIG. 15. The mean-field theory Curie temperatures ($T_C^{MFT}$) of FeB alloys with Ti ($Fe_{1-x}Ti_xB$) calculated with FPLO5 code and Perdew and Wang (PW92) functional. The chemical disorder was modeled with coherent potential approximation (CPA) and the paramagnetic state with disorder local moment (DLM-CPA) method. For $\beta$-FeB phase, the experimental Curie temperature is around 590 K [21].

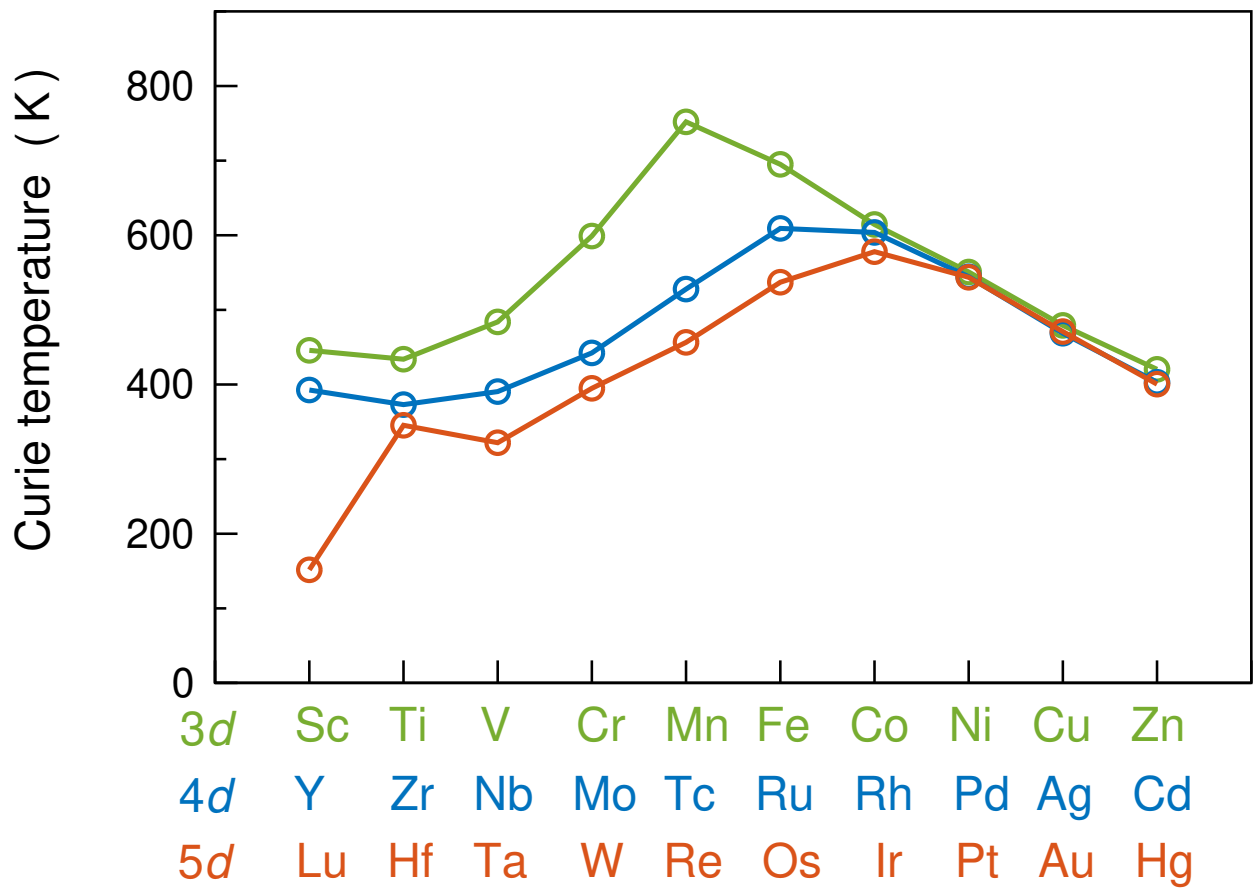

FIG. 16. The mean-field theory Curie temperatures ($T_C^{MFT}$) of $Fe_{0.917}X_{0.083}B$ alloys with $3d$, $4d$, and $5d$ transition metals X calculated with FPLO5 code and Perdew and Wang (PW92) functional. The chemical disorder was modeled with coherent potential approximation (CPA) and the paramagnetic state with disorder local moment (DLM-CPA) method. For $\beta$-FeB phase, the experimental Curie temperature is around 590 K [21].

netic hardnesses of FeB alloys appear more promising.

We will start non-sequentially with Fig. 12(c) showing the orbital magnetic moments on dopant sites. The trends resemble very much the ones for MnB alloys presented above. However, the orbital moments of FeB alloys are slightly lower than those of MnB alloys. The value of about 0.03 $\mu_B$ on Fe in FeB is similar to the value of 0.043 $\mu_B$ calculated for bcc Fe. Both values, however, are significantly underestimated relative to the experimental orbital moment in bcc Fe equal to 0.086 $\mu_B$ [75]. The observed dependencies of the orbital moment on the atomic number of the $5d$ elements resemble the earlier theoretical results for $5d$ dopants in Fe [79], $CeFe_{12}$ [67], and $Fe_5PB_2$ [48].

In Fig. 12(b), we observe the spin polarization of dopants in a magnetic medium of FeB. The dependencies look similar to the results for MnB alloys. Related trends in spin magnetic moments for the transition metal elements in the $3d$ magnetic media (Fe and Ni) have been calculated before [79, 80]. Similar magnetic behavior of $5d$ dopants in Fe has also been measured by spin-dependent absorption [81].

In Fig. 12(a) we see the total spin magnetic moments of $Fe_{11}X_1B_{12}$ monoborides. Alloying of FeB with transition metals from periodic groups above the Fe group systematically decreases the total moment. In contrast, although several elements in the groups preceding Fe continue the trend of increasing total moment, it quickly collapses, and does not pursue values close to the 2.0 $\mu_B$ f.u.$^{-1}$ observed for MnB. Among the $Fe_{11}X_1B_{12}$ alloys, the FeB alloy with Cr has the highest total spin magnetic moment of 1.34 $\mu_B$ f.u.$^{-1}$, compared to 1.22 $\mu_B$ f.u.$^{-1}$ for FeB alone.

The direction of the magnetic easy axis for FeB is [010] and in most cases, substituting FeB with transi-tion metals does not change the easy axis, see Table IV. Figures 12(d) and 12(e) present the magnetocrystalline anisotropy energies and corresponding magnetic hardnesses. The MAE clearly increases for substitutions from groups preceding Fe, while it remains at a relatively low level for substitutions from groups succeeding Fe. Such sufficiently high MAE values, in combination with low magnetic moments, lead to a magnetic hardness above unity, i.e. qualify these alloys as magnetically hard. $Fe_{11}Ti_1B_{12}$ has the highest magnetic hardness equal to 1.15. Thus, it is a prospective candidate for permanent magnets, together with other FeB alloys with Sc, V, Zr, Nb, Mo, Ta or W (all indicating magnetic hardness above unity). In previous work, we predicted that similar transition metal substitutions increases the magnetic hardness of $CeFe_{12}$-based alloys [67].

In the case of the considered monoborides with orthorhombic structure, we defined the MAE as the energy difference between two lower total energies among three total energies calculated for magnetization along three main crystallographic axes. The energy difference between two higher energies ($DE_{32}$) complements the MAE. For the listed above $Fe_{11}X_1B_{12}$ alloys with $\kappa > 1$, the $DE_{32}$ is small (about 0.1 $MJ\,m^{-3}$), indicating uniaxial-like anisotropy. For $Fe_{11}Os_1B_{12}$, the $DE_{32}$ is about 2.0 $MJ\,m^{-3}$, which together with a MAE close to zero, implies anisotropy in type of easy plane. In contrast, the magnetic anisotropy of systems with intermediate MAE and $DE_{32}$ values is characteristic for orthorhombic systems.

As we did above for $Fe_{0.5}Co_{0.5}B$ and FeB with fixed spin moment, here we will discuss aspects of the band structure of FeB alloys with transition metals that affect the MAE. For this purpose, we choose the FeB alloy with Ti ($Fe_{11}Ti_1B_{12}$) as an example of an alloy with a relatively high MAE (0.86 $MJ\,m^{-3}$) and compare its band structure with that of the initial FeB compound (MAE = 0.19 $MJ\,m^{-3}$). In Fig. 13, we see the two mentioned band structures in a range from -2 to 2 eV. The valence band is dominated by contributions from 3*d* orbitals. Replacing one of the twelve Fe atoms in the supercell with a Ti atom, induces a reduction in symmetry and leads to seven non-equivalent Fe atoms in the cell. In the band structure image, this leads to multiple splitting of bands originating from Fe atoms and the appearance of additional bands coming from the Ti atom. The occupancy of the Fe 3*d* and Ti 3*d* bands is about 6.5 and 2.5, respectively, making the Ti 3*d* bands shift upward on the energy scale and lowering the Fermi level slightly. Moreover, the weighted contributions from Ti 3*d* states show the presence of Ti 3*d* majority states near the Fermi level, see Fig. 13(c). To sum up, it is difficult to identify a single cause for the favorable effect of Ti on the MAE of the alloy. The significant increase in MAE after alloying FeB with Ti is the result of the constructive sum of effects from split Fe 3*d* bands, shift of the Fermi level, additional bands from Ti, and modification of spin polarization.

Total spin magnetic moments presented in Fig. 12(a) show a somewhat unexpected bump for the Ti and V columns. To investigate this behavior, in Fig. 14(a) we have juxtaposed the results of 3*d*-substituted supercells with those from CPA. In the CPA, all magnetic moments are lower, and alloys with Ti and V show no additional peak. However, since the CPA results come from LDA, we cannot determine whether the discrepancy comes from the choice of chemical disorder modeling method (supercells vs. CPA) or exchange-correlation potential (GGA vs. LDA). For this reason, we performed also supercells-LDA calculations, and in these results we also see a clear bump in the moments on Ti and V. We conclude that the supercell method is responsible for the boost in moments on Ti and V, and alloying FeB with Ti and V will most likely not lead to an increase in magnetic moment. However, all three approaches considered predict an increase in the total moment of FeB alloys after substituting with Mn and Cr. A comparison of magnetic moments on transition metal atoms from two LDA approaches (supercell and CPA), see Fig. 14(b), besides noticeable differences for substituted Ti, V, and Cr shows also and nearly identical results for elements from Mn upwards.

When considering materials for use as permanent magnets, one of the basic parameters that should be taken into account is the Curie temperature, which must be significantly higher than room temperature. The experimental Curie temperature of the FeB is around 590 K [21]. It is therefore necessary to ask what effect the substitutions under consideration will have on it. While we know that the addition of Mn has a positive effect, we expect all other transition metals to lower $T_C$. The mean-field theory Curie temperatures ($T_C^{MFT}$) we calculated for FeB is 695 K, see Eq. 4. It overestimates the experimental value by nearly twenty percent. We assume that the $T_C$s calculated here for all alloys are overestimated similarly, and we will focus on the relationships between the temperatures determined for the initial FeB and doped systems. Calculations for Ti substituting confirm the expected decrease in $T_C$ with increasing Ti concentration in the alloy, see Fig. 15. With the addition of about 8% Ti reduces $T_C$ by almost 40%. In Fig. 16 we present the $T_C^{MFT}$ calculated for $Fe_{0.917}X_{0.083}B$ alloys. As expected, all dopants, except Mn, lower the $T_C$ of the initial FeB compound. The trends observed along the three periods are characterized by clear maxima in the central part. From the perspective of the material's applications as a permanent magnet and the preservation of as high a Curie temperature as possible, it is better to use elements from the middle of the periods as substitutions. In our case, the optimal candidate may be Cr, which leads to a significant increase in MAE, see Fig. 12(d), with the least possible impact on $T_C$.

In conclusion, from the point of view of application in permanent magnets, considering the results of calculations of formation energy, Curie temperature, and magnetocrystalline anisotropy energy, the most promising are FeB alloys with Mn and Cr. They should be stable and have the highest possible Curie temperatures. The hypothetical maximum MAE at high temperatures we estimated only indirectly from the fixed spin moment calculations (at 0 K) for FeB as about 1.5 MJ m$^{-3}$, giving a magnetic hardness above 1.0. (A magnetic hardness of 0.8 for pure FeB has been obtained experimentally.) Although for transition metals we considered substitution concentrations around 8%, it is also worth experimentally testing the MAE temperature dependence of both lower and higher concentrations to find the optimal composition. Finally, to translate an intrinsic material quantity such as MAE into coercivity it will be necessary to further optimize the microstructure of the alloys.

## IV. SUMMARY AND CONCLUSIONS

Magnetic monoborides have sufficiently high values of magnetization and Curie temperature that some of them can be considered for permanent magnet applications. However, their main representatives, which are MnB and FeB, show very low values of coercivity. Probably for this reason, there has been almost no studies on the magnetocrystalline anisotropy of magnetic monoborides to date. The only exception is an experiment on FeB monocrystals, which, surprisingly, classified it as semi-hard magnetic and close to magnetically hard material.

In this work, by performing first-principles calculations, we have taken a step towards determining the magnetic anisotropic properties of magnetic monoborides. We considered the entire range of monoboride solid solutions from CrB, through MnB, FeB, to CoB and, plus a whole series of MnB and FeB alloys substituted with the transition metal elements 3*d*, 4*d* and 5*d*. We found that in the concentration range from CrB to CoB, the calculated magnetic easy axis changes direction several times. In the range between MnB and FeB, the determined magnetic hardness approaches unity three times, whereby exceeding unity classifies the material as magnetically hard. In the range between FeB and CoB, the magnetic hardness increases with the Co concentration to exceed an impressive five in the middle of the range. Although the Curie temperature is there close to room temperature, we expect that high magnetic hardness will also occur at lower Co concentrations, showing somewhat higher Curie temperatures.

For alloys of MnB and FeB with transition metals, we have identified characteristic trends in magnetic moments and magnetic hardness along the transition metal periods. In the case of FeB alloys, most transition metals from groups preceding Fe increase the magnetic hardness of the alloys. While the magnetic hardness of MnB alloys remains low or medium, for several FeB alloys it exceeds unity. As magnetically hard, we have classified FeB alloys ($Fe_{11}X_1B_{12}$) with Sc, Ti, V, Zr, Nb, Mo, Hf, Ta or W.

Calculations of formation energies imply potential stability of all considered alloys with transition metals. Calculations of Curie temperatures showed that all substitutions, except Mn, significantly lower the Curie temperature of FeB alloys, with the smallest effect observed around the middle of the periods.

The presented calculation results encourage their verification by experimental methods, with additional extension of the measurements to include the temperature dependence of magnetic anisotropy.

## ACKNOWLEDGMENTS

We gratefully acknowledge financial support from the National Science Center Poland under decisions DEC-2019/35/O/ST5/02980 (PRELUDIUM-BIS 1) and DEC-2021/41/B/ST5/02894 (OPUS 21). We thank Zbigniew Śniadecki and Justyna Rychły-Gruszecka for valuable comments. We thank Paweł Leśniak and Daniel Depcik for compiling the scientific software and administering the computational cluster at the Institute of Molecular Physics, Polish Academy of Sciences. Part of the calculation was made in the Poznan Supercomputing and Networking Centre (PSNC/PCSS).

[1] T. Kanaizuka, Invar like properties of transition metal monoborides $Mn_{1-x}Cr_xB$ and $Mn_{1-x}Fe_xB$, Mater. Res. Bull. **16**, 1601 (1981).

[2] O. V. Zhdanova, M. B. Lyakhova, and Yu. G. Pastushenkov, Magnetic properties and domain structure of FeB single crystals, Met. Sci. Heat Treat. **55**, 68 (2013).

[3] A. Edström, M. Werwiński, D. Iuşan, J. Rusz, O. Eriksson, K. P. Skokov, I. A. Radulov, S. Ener, M. D. Kuz'min, J. Hong, M. Fries, D. Yu. Karpenkov, O. Gutfleisch, P. Toson, and J. Fidler, Magnetic properties of $(Fe_{1-x}Co_x)_2B$ alloys and the effect of doping by 5d elements, Phys. Rev. B **92**, 174413 (2015).

[4] T. N. Lamichhane, O. Palasyuk, V. P. Antropov, I. A. Zhuravlev, K. D. Belashchenko, I. C. Nlebedim, K. W. Dennis, A. Jesche, M. J. Kramer, S. L. Bud'ko, R. W. McCallum, P. C. Canfield, and V. Taufour, Reinvestigation of the intrinsic magnetic properties of $(Fe_{1-x}Co_x)_2B$ alloys and crystallization behavior of ribbons, J. Magn. Magn. Mater. **513**, 167214 (2020).

[5] N. Lundquist, H. P. Myers, and R. Westin, The paramagnetic properties of the monoborides of V, Cr, Mn, Fe, Co and Ni, Philos. Mag. **7**, 1187 (1962).

[6] M. C. Cadeville and E. Daniel, Sur la structure électronique de quelques borures d'éléments de transition, J. Phys. **27**, 449 (1966).

[7] S. Carenco, D. Portehault, C. Boissière, N. Mézailles, and C. Sanchez, Nanoscaled Metal Borides and Phosphides: Recent Developments and Perspectives, Chem. Rev. **113**, 7981 (2013).

[8] Z. Pu, T. Liu, G. Zhang, X. Liu, Marc. A. Gauthier, Z. Chen, and S. Sun, Nanostructured Metal Borides for Energy-Related Electrocatalysis: Recent Progress, Challenges, and Perspectives, Small Methods **5**, 2100699 (2021).

[9] M. Fries, Z. Gercsi, S. Ener, K. P. Skokov, and O. Gutfleisch, Magnetic, magnetocaloric and structural properties of manganese based monoborides doped with iron and cobalt – A candidate for thermomagnetic generators, Acta Mater. **113**, 213 (2016).

[10] S. Ma, K. Bao, Q. Tao, P. Zhu, T. Ma, B. Liu, Y. Liu, and T. Cui, Manganese mono-boride, an inexpensive room temperature ferromagnetic hard material, Sci. Rep. **7**, 43759 (2017).

[11] J. D. Bocarsly, E. E. Levin, S. A. Humphrey, T. Faske, W. Donner, S. D. Wilson, and R. Seshadri, Magnetostructural Coupling Drives Magnetocaloric Behavior: The Case of MnB versus FeB, Chem. Mater. **31**, 4873 (2019).

[12] H. M. Sánchez, L. E. Z. Alfonso, J. S. T. Hernandez, D. Salazar, and G. A. P. Alcázar, Improving the ferromagnetic exchange coupling in hard $\tau$-$Mn_{53.3}Al_{45.0}C_{1.7}$ and soft $Mn_{50}B_{50}$ magnetic alloys, Appl. Phys. A **126**, 843 (2020).

[13] S. Klemenz, M. Fries, M. Dürrschnabel, K. Skokov, H.-J. Kleebe, O. Gutfleisch, and B. Albert, Low-temperature synthesis of nanoscale ferromagnetic $\alpha$'-MnB, Dalton Trans. **49**, 131 (2020).

[14] S. Ma, R. Farla, K. Bao, A. Tayal, Y. Zhao, Q. Tao, X. Yang, T. Ma, P. Zhu, and T. Cui, An electrically conductive and ferromagnetic nano-structure manganese mono-boride with high Vickers hardness, Nanoscale **13**, 18570 (2021).

[15] N. Kalyon, A.-M. Zieschang, K. Hofmann, M. Lepple, M. Fries, K. P. Skokov, M. Dürrschnabel, H.-J. Kleebe, O. Gutfleisch, and B. Albert, CrB-type, ordered $\alpha$ -MnB: Single crystal structure and spin-canted magnetic behavior, APL Mater. **11**, 060701 (2023).

[16] A. N. Kolmogorov, S. Shah, E. R. Margine, A. F. Bialon, T. Hammerschmidt, and R. Drautz, New Superconducting and Semiconducting Fe-B Compounds Predicted with an *Ab Initio* Evolutionary Search, Phys. Rev. Lett. **105**, 217003 (2010).

[17] L. Han, S. Wang, J. Zhu, S. Han, W. Li, B. Chen, X. Wang, X. Yu, B. Liu, R. Zhang, Y. Long, J. Cheng, J. Zhang, Y. Zhao, and C. Jin, Hardness, elastic, and electronic properties of chromium monoboride, Appl. Phys. Lett. **106**, 221902 (2015).

[18] T. Lundstrom, Structure, defects and properties of some refractory borides, Pure Appl. Chem. **57**, 1383 (1985).

[19] C. Kapfenberger, B. Albert, R. Pöttgen, and H. Huppertz, Structure refinements of iron borides $Fe_2B$ and FeB, Z. Krist. **221**, 477 (2006).

[20] F. Igoa Saldaña, E. Defoy, D. Janisch, G. Rousse, P.-O. Autran, A. Ghoridi, A. Séné, M. Baron, L. Suescun, Y. Le Godec, and D. Portehault, Revealing the Elusive Structure and Reactivity of Iron Boride $\alpha$-FeB, Inorg. Chem. **62**, 2073 (2023).

[21] S. Rades, S. Kraemer, R. Seshadri, and B. Albert, Size and Crystallinity Dependence of Magnetism in Nanoscale Iron Boride, $\alpha$-FeB, Chem. Mater. **26**, 1549 (2014).

[22] T. Kanaizuka, Phase diagram of pseudobinary CrB-MnB and MnB-FeB systems: Crystal structure of the low-temperature modification of FeB, J. Solid State Chem. **41**, 195 (1982).

[23] Y. Wang, L. Li, Y. Wang, D. Song, G. Liu, Y. Han, L. Jiao, and H. Yuan, Crystalline CoB: Solid state reaction synthesis and electrochemical properties, J. Power Sources **196**, 5731 (2011).

[24] Z. Guo, L. Zhang, T. Azam, and Z.-S. Wu, Recent advances and key challenges of the emerging MBenes from synthesis to applications, MetalMat , e12 (2023).

[25] H. Zhang, H. Xiang, F.-z. Dai, Z. Zhang, and Y. Zhou, First demonstration of possible two-dimensional MBene CrB derived from MAB phase $Cr_2AlB_2$, J. Mater. Sci. Technol. **34**, 2022 (2018).

[26] Y. Wang, W. Xu, D. Yang, Y. Zhang, Y. Xu, Z. Cheng, X. Mi, Y. Wu, Y. Liu, Y. Hao, and G.-Q. Han, Above-Room-Temperature Strong Ferromagnetism in 2D MnB Nanosheet, ACS Nano **17**, 24320 (2023).

[27] J.-l. Ma, N. Li, Q. Zhang, X.-b. Zhang, J. Wang, K. Li, X.-f. Hao, and J.-m. Yan, Synthesis of porous and metallic CoB nanosheets towards a highly efficient electrocatalyst for rechargeable Na–$O_2$ batteries, Energy Environ. Sci. **11**, 2833 (2018).

[28] M. Dou, H. Li, Q. Yao, J. Wang, Y. Liu, and F. Wu, Room-temperature ferromagnetism in two-dimensional transition metal borides: A first-principles investigation, Phys. Chem. Chem. Phys. **23**, 10615 (2021).

[29] S. Wang, N. Miao, K. Su, V. A. Blatov, and J. Wang, Discovery of intrinsic two-dimensional antiferromagnets from transition-metal borides, Nanoscale **13**, 8254 (2021).

[30] Z. Jiang, P. Wang, X. Jiang, and J. Zhao, MBene (MnB): A new type of 2D metallic ferromagnet with high Curie temperature, Nanoscale Horiz. **3**, 335 (2018).

[31] C. Liu, B. Fu, H. Yin, G. Zhang, and C. Dong, Strain-tunable magnetism and nodal loops in monolayer MnB, Appl. Phys. Lett. **117**, 103101 (2020).

[32] J. Tang, S. Li, D. Wang, Q. Zheng, J. Zhang, T. Lu, J. Yu, L. Sun, B. Sa, B. G. Sumpter, J. Huang, and W. Sun, Enriching 2D transition metal borides *via* MB XMenes (M = Fe, Co, Ir): Strong correlation and magnetism, Nanoscale Horiz. **9**, 162 (2024).

[33] B. Wang, D. Y. Wang, Z. Cheng, X. Wang, and Y. X. Wang, Phase Stability and Elastic Properties of Chromium Borides with Various Stoichiometries, ChemPhysChem **14**, 1245 (2013).

[34] X. Xu, K. Fu, L. Li, Z. Lu, X. Zhang, Y. Fan, J. Lin, G. Liu, H. Luo, and C. Tang, Dependence of the elastic properties of the early-transition-metal monoborides on their electronic structures: A density functional theory study, Phys. B Condens. Matter **419**, 105 (2013).

[35] B. Wang, X. Li, Y. X. Wang, and Y. F. Tu, Phase Stability and Physical Properties of Manganese Borides: A First-Principles Study, J. Phys. Chem. C **115**, 21429 (2011).

[36] J. Park, Y.-K. Hong, H.-K. Kim, W. Lee, C.-D. Yeo, S.-G. Kim, M.-H. Jung, C.-J. Choi, and O. N. Mryasov, Electronic structures of MnB soft magnet, AIP Adv. **6**, 055911 (2016).

[37] J. Hafner, M. Tegze, and Ch. Becker, Amorphous magnetism in Fe-B alloys: First-principles spin-polarized electronic-structure calculations, Phys. Rev. B **49**, 285 (1994).

[38] P. Mohn and D. G. Pettifor, The calculated electronic and structural properties of the transition-metal monoborides, J. Phys. C Solid State Phys. **21**, 2829 (1988).

[39] Y. Bourourou, L. Beldi, B. Bentria, A. Gueddouh, and B. Bouhafs, Structure and magnetic properties of the 3d transition-metal mono-borides TM–B (TM=Mn, Fe, Co) under pressures, J. Magn. Magn. Mater. **365**, 23 (2014).

[40] A. Gueddouh, The effects of magnetic moment collapse under high pressure, on physical properties in mono-borides TMB (TM = Mn, Fe): A first-principles,

Phase Transit. **90**, 984 (2017).

[41] S. Nishiyama, H. Nakamura, and T. Hattori, Thermoelectric Properties and Electronic Density of States of Transition Metal Monoborides, FeB, CoB, NiB and their Solid Solutions, J. Ceram. Soc. Jpn. Suppl. **112**, S676 (2004).

[42] S. Nakamura and Y. Gohda, Prediction of ferromagnetism in MnB and MnC on nonmagnetic transition-metal surfaces studied by first-principles calculations, Phys. Rev. B **96**, 245416 (2017).

[43] X. Tan, Z. Na, R. Zhuo, D. Wang, Y. Zhang, and P. Wu, Investigation of CrB as a Potential Gas Sensor for Fault Detection in Eco-Friendly Power Equipment, Chemosensors **11**, 371 (2023).

[44] C. Romero-Muñiz, J. Y. Law, L. M. Moreno-Ramírez, Á. Díaz-García, and V. Franco, Using a computationally driven screening to enhance magnetocaloric effect of metal monoborides, J. Phys. Energy **5**, 024021 (2023).

[45] R. Kadrekar, N. Patel, and A. Arya, Understanding the role of boron and stoichiometric ratio in the catalytic performance of amorphous Co-B catalyst, Appl. Surf. Sci. **518**, 146199 (2020).

[46] H. Zhang, B. Zhao, F.-Z. Dai, H. Xiang, Z. Zhang, and Y. Zhou, $(Cr_{0.2}Mn_{0.2}Fe_{0.2}Co_{0.2}Mo_{0.2})B$: A novel high-entropy monoboride with good electromagnetic interference shielding performance in K-band, J. Mater. Sci. Technol. **77**, 58 (2021).

[47] D. Hedlund, J. Cedervall, A. Edström, M. Werwiński, S. Kontos, O. Eriksson, J. Rusz, P. Svedlindh, M. Sahlberg, and K. Gunnarsson, Magnetic properties of the $Fe_5SiB_2$ - $Fe_5PB_2$ system, Phys. Rev. B **96**, 094433 (2017).

[48] M. Werwiński, A. Edström, J. Rusz, D. Hedlund, K. Gunnarsson, P. Svedlindh, J. Cedervall, and M. Sahlberg, Magnetocrystalline anisotropy of $Fe_5PB_2$ and its alloys with Co and 5d elements: A combined first-principles and experimental study, Phys. Rev. B **98**, 214431 (2018).

[49] J. Cedervall, E. Nonnet, D. Hedlund, L. Häggström, T. Ericsson, M. Werwiński, A. Edström, J. Rusz, P. Svedlindh, K. Gunnarsson, and M. Sahlberg, Influence of Cobalt Substitution on the Magnetic Properties of $Fe_5PB_2$, Inorg. Chem. **57**, 777 (2018).

[50] S. Pal, K. Skokov, T. Groeb, S. Ener, and O. Gutfleisch, Properties of magnetically semi-hard $(Fe_xCo_{1-x})_3B$ compounds, J. Alloys Compd. **696**, 543 (2017).

[51] M. Däne, S. K. Kim, M. P. Surh, D. Åberg, and L. X. Benedict, Density functional theory calculations of magnetocrystalline anisotropy energies for $(Fe_{1-x}Co_x)_2B$, J. Phys.: Condens. Matter **27**, 266002 (2015).

[52] K. Bourzac, The rare-earth crisis, Technol. Rev. **114**, 58 (2011).

[53] J. Coey, Permanent magnets: Plugging the gap, Scr. Mater. **67**, 524 (2012).

[54] R. Skomski and J. Coey, Magnetic anisotropy — How much is enough for a permanent magnet?, Scr. Mater. **112**, 3 (2016).

[55] K. Koepernik and H. Eschrig, Full-potential nonorthogonal local-orbital minimum-basis band-structure scheme, Phys. Rev. B **59**, 1743 (1999).

[56] I. Opahle, K. Koepernik, and H. Eschrig, Full-potential band-structure calculation of iron pyrite, Phys. Rev. B **60**, 14035 (1999).

[57] J. P. Perdew, K. Burke, and M. Ernzerhof, Generalized gradient approximation made simple, Phys. Rev. Lett. **77**, 3865 (1996).

[58] U. von Barth and L. Hedin, A local exchange-correlation potential for the spin polarized case. i, J. Phys. C Solid State Phys. **5**, 1629 (1972).

[59] J. P. Perdew and Y. Wang, Accurate and simple analytic representation of the electron-gas correlation energy, Phys. Rev. B **45**, 13244 (1992).

[60] K. Momma and F. Izumi, *VESTA* : A three-dimensional visualization system for electronic and structural analysis, J. Appl. Crystallogr. **41**, 653 (2008).

[61] A. I. Liechtenstein, M. I. Katsnelson, V. P. Antropov, and V. A. Gubanov, Local spin density functional approach to the theory of exchange interactions in ferromagnetic metals and alloys, J. Magn. Magn. Mater. **67**, 65 (1987).

[62] R. Wu and A. J. Freeman, Spin–orbit induced magnetic phenomena in bulk metals and their surfaces and interfaces, J. Magn. Magn. Mater. **200**, 498 (1999).

[63] B. L. Gyorffy, A. J. Pindor, J. Staunton, G. M. Stocks, and H. Winter, A first-principles theory of ferromagnetic phase transitions in metals, J. Phys. F Met. Phys. **15**, 1337 (1985).

[64] J. Kudrnovský, I. Turek, V. Drchal, F. Máca, P. Weinberger, and P. Bruno, Exchange interactions in III-V and group-IV diluted magnetic semiconductors, Phys. Rev. B **69**, 115208 (2004).

[65] V. Heine, J. H. Samson, and C. M. M. Nex, Theory of local magnetic moments in transition metals, J. Phys. F Met. Phys. **11**, 2645 (1981).

[66] P. Soven, Coherent-potential model of substitutional disordered alloys, Phys. Rev. **156**, 809 (1967).

[67] J. Snarski-Adamski and M. Werwiński, Effect of transition metal doping on magnetic hardness of $CeFe_{12}$-based compounds, J. Magn. Magn. Mater. **554**, 169309 (2022).

[68] R. Kuentzler, Specific Heat of Nearly Ferromagnetic Borides $(Co–Ni)_2B$ and (Fe–Co)B, J. Appl. Phys. **41**, 908 (1970).

[69] L. Takacs, M. C. Cadeville, and I. Vincze, Mossbauer study of the intermetallic compounds $(Fe_{1-x}Co_x)_2B$ and $(Fe_{1-x}Co_x)B$, J. Phys. F Met. Phys. **5**, 800 (1975).

[70] P.-H. Lee, S.-H. Chen, Y.-A. Chen, K.-L. Chen, T.-W. Wang, S.-W. Wang, B.-S. Wu, W.-S. Su, and S.-K. Chen, Calculated Magnetism on $Mn_{1-x}Fe_xB$ Alloys, BAOJ Phys. **2**, 007 (2016).

[71] A. Gueddouh, A. Benghia, and S. Maabed, Effect of Mn content in $Fe_{(1-x)}Mn_xB$ (x = 0, 0.25, 0.5, 0.75 and 1) on physical properties - ab initio calculations, Mater. Sci.-Pol. **37**, 71 (2019).

[72] N. A. Klindukhov, V. S. Kasperovich, M. G. Shelyapina, and H. El Kebir, Calculation of the electronic structure and hyperfine fields for $Fe_{1-x}Co_xB$ and $(Fe_{1-x}Co_x)_2B$ compounds by the Korringa-Kohn-Rostoker method, Phys. Solid State **50**, 302 (2008).

[73] P. Lee, Z. Xiao, K. Chen, Y. Chen, S. Kao, and T. Chin, The magnetism of $Fe_{(1-x)}Co_xB$ alloys: First principle calculations, Phys. B Condens. Matter **404**, 1989 (2009).

[74] P. Bruno, Tight-binding approach to the orbital magnetic moment and magnetocrystalline anisotropy of transition-metal monolayers, Phys. Rev. B **39**, 865 (1989).

[75] C. T. Chen, Y. U. Idzerda, H.-J. Lin, N. V. Smith, G. Meigs, E. Chaban, G. H. Ho, E. Pellegrin, and F. Sette, Experimental Confirmation of the X-Ray Magnetic Circular Dichroism Sum Rules for Iron and Cobalt, Phys. Rev. Lett. **75**, 152 (1995).

[76] R. S. Mulliken, Electronic Population Analy-

sis on LCAO–MO Molecular Wave Functions. I, J. Chem. Phys. **23**, 1833 (1955).

[77] T. Burkert, L. Nordström, O. Eriksson, and O. Heinonen, Giant Magnetic Anisotropy in Tetragonal FeCo Alloys, Phys. Rev. Lett. **93**, 027203 (2004).

[78] K. D. Belashchenko, L. Ke, M. Däne, L. X. Benedict, T. N. Lamichhane, V. Taufour, A. Jesche, S. L. Bud'ko, P. C. Canfield, and V. P. Antropov, Origin of the spin reorientation transitions in $(Fe_{1-x}Co_x)_2B$ alloys, Appl. Phys. Lett. **106**, 062408 (2015).

[79] P. H. Dederichs, R. Zeller, H. Akai, and H. Ebert, Ab-initio calculations of the electronic structure of impurities and alloys of ferromagnetic transition metals, J. Magn. Magn. Mater. **100**, 241 (1991).

[80] H. Akai, Nuclear spin-lattice relaxation of impurities in ferromagnetic iron, Hyperfine Interact. **43**, 253 (1988).

[81] R. Wienke, G. Schütz, and H. Ebert, Determination of local magnetic moments of 5d impurities in Fe detected via spin-dependent absorption, J. Appl. Phys. **69**, 6147 (1991).